\documentclass[a4paper]{jpconf}

\usepackage{graphicx} 

\newcommand{\bra}[1]{\langle #1|}
\newcommand{\ket}[1]{|#1\rangle}
\newcommand{\braket}[2]{\langle #1|#2\rangle}

\begin{document}

\title{Entanglement dynamics in a quantum-classical hybrid of two q-bits and one oscillator}  

\author{L. Fratino$^a$, A. Lampo$^b$  and H.-T. Elze$^{c,}$\footnote{Corresponding author.}}

\address{$^a$ SEPnet \& Hubbard Theory Consortium, Department of Physics, Royal Holloway, 
              University of London, Egham, Surrey TW20 0EX, UK}

\address{$^b$ ICFO, Mediterranean Technology Park, Av. C.F. Gauss 3, 08860 Castelldefels 
              (Barcelona), Spain} 

\address{$^c$ Dipartimento di Fisica ``Enrico Fermi'',  
              Largo Pontecorvo 3, 56127 Pisa, Italia }

\ead{lorenzo.fratino.2013@live.rhul.ac.uk, aniello.lampo@icfo.es, elze@df.unipi.it}

\begin{abstract} 
We investigate new features, especially of entanglement dynamics, which arise in a 
quantum-classical hybrid. As a model, we study the coupling between two quantum mechanical 
two-level systems, i.e. two q-bits, and a classical harmonic oscillator. Their 
interaction is described by a hybrid coupling, in accordance with a 
recently developed quantum-classical hybrid theory. We discuss various situations
in which entanglement of the q-bits does (not) evolve. Furthermore, we point out 
an experimental application in a hybrid cooling scheme and indicate topics for future study.

\end{abstract}

\section{Introduction}
Microscopic and macroscopic objects are commonly described by means of two different kinds of theories: 
the former through quantum mechanics and the latter by classical mechanics. Despite its successes, 
this ``brutal'' division confines these two different kinds of objects in two noncommunicating ``worlds'', 
the boundaries of which are still not clearly delineated. In fact, it has been unknown since the inception of 
quantum theory whether a borderline between the classical and quantum worlds exists or not. 
In order to unify quantum and classical descriptions, we make use of a recent 
generalization of these theories called {\it quantum-classical hybrid theory} \cite{elze1,buric1,buric2,hall1}.
This approach builds a bridge between classical and quantum mechanics, proposing a consistent mathematical framework 
for the description of a quantum-classical mixed reality and a tool for future applications, e.g., in measurement 
or quantum information protocols that incorporate classical control of or read-out from a quantum object \cite{Diosi}. 
Hybrid theory and its applications will be relevant here, since laboratory quantum devices, at some stage, 
incorporate a quantum-classical interface.

Presently, we study one of the most peculiar properties of multipartite quantum states, their {\it entanglement}, 
in situations involving quantum-classical hybrid systems.

A state is {\it entangled}, if and only if it is not separable into states that describe independent subsystems 
of a composite. A composite system is disentangled or separable, in turn, if and only if its 
state can be factorized. 
The main question that motivated this work is: What kind of new features arise in this 
context in the presence of a quantum-classical hybrid coupling? 
In order to answer this, we consider a model that describes the coupling between two 
{\it quantum-mechanical two-level (spin) systems}, i.e. two q-bits, and a {\it classical harmonic oscillator}.
 
We base our analysis on the quantum-classical hybrid approach developed in Refs.\,\cite{elze1,elze2,elze3}, 
which satisfies a large list of consistency requirements discussed in the recent literature, as summarized in 
Ref.\,\cite{elze4} . 

The model of a harmonic oscillator coupled to two q-bits, when treated fully quantum mechanically, exhibits 
a great richness of entanglement dynamics, see, for example, Refs.\,\cite{atomcavity,photocavity,irish1,irish2,tavis2}.   
Presently, we consider the case that the oscillator part of this composite system is strictly classical and, therefore, its 
coupling to the q-bits must be of hybrid nature.  

The Hamiltonian of this system consists of a classical, a quantum mechanical, and a hybrid part, respectively (choosing units such that 
$\hbar =1$):
\begin{eqnarray}\label{energycl}
 H_{cl}&:=&\frac{p^2}{2m}+\frac{m\omega^2 x^2}{2}\;\;, \\ [1ex] \label{hqm}
 \hat H_{qm}&:=&\omega_0\hat S_z\;\;, \\ [1ex] \label{hhybrid}
 \hat\mathcal{I}_{hyb}&:=&\beta\omega\sqrt{m\omega}\; x \hat S_x=\beta\Omega\; x \hat S_x \;\;,
\end{eqnarray}
where the two-q-bit (``$A$'' and ``$B$'') operators are $\hat{S}_x\equiv (\hat{\sigma}^A_x+\hat{\sigma}^B_x)/2$ and $\hat{S}_z\equiv (\hat{\sigma}^A_z+\hat{\sigma}^B_z)/2$, 
and where we introduced $\Omega\equiv\omega\sqrt{m\omega}$; $\beta$ is a dimensionless constant determining the coupling strength, 
which can be expressed in terms of microscopic physical parameters depending on the physical realization of the model (cf. Section\,2). We recall that, 
with appropriate substitutions, the contribution $H_{cl}$, i.e. the classical Hamiltonian, can describe a single mode of  a classical electric field.

Following hybrid theory \cite{elze1}, we proceed to construct the full {\it hybrid Hamiltonian function} from the above contributions. In particular, 
introducing the {\it oscillator expansion} of a generic state $\ket{\psi}$ in the eigenbasis of $\hat{S}_x$ \cite{elze2,heslot}:
\begin{equation} \label{oscexp}
 \ket{\psi}=\sum_{\alpha=1}^4\frac{X_\alpha+\imath P_\alpha}{\sqrt{2}}\ket{\phi_\alpha} \;\;,
\end{equation}
where the $\ket{\phi_\alpha}$ denote, respectively, the states $\ket{++}$, $\ket{--}$, $(\ket{+-}+\ket{-+})/\sqrt2$, $(\ket{+-}-\ket{-+})/\sqrt2$, 
with $\hat\sigma_x\ket{\pm}=\pm\ket{\pm}$. The ``coordinates'' and ``momenta'' introduced here have to satisfy a constraint, which 
stems from the wave function normalization, $\langle\psi(t)|\psi(t)\rangle=1$, i.e.:
\begin{equation}\label{constraint}
\mathcal{C}(X_\alpha,P_\alpha ):=\sum_{\alpha=1}^4(X^2_\alpha+P_\alpha^2)=2 \;\;.
\end{equation}
Based on this expansion, contributions to the full Hamiltonian function from eqs.\,(\ref{hqm})--(\ref{hhybrid}) become:
\begin{eqnarray}\label{hqm1}
 H_{qm}&:=&\bra{\psi}\hat H_{qm}\ket{\psi}=\frac{\omega_0}{2}\sum_{\alpha,\beta=1}^4(X_\alpha+\imath P_\alpha)S_{\alpha \beta}(X_\beta-\imath P_\beta) 
\;\;, \\ [1ex] \label{hhybrid1}  
 \mathcal{I}_{hyb}&:=&\bra{\psi}\hat\mathcal{I}_{hyb}\ket{\psi}=\frac{\beta\Omega}{2} x\sum_{\alpha=1}^4 E_\alpha (X_\alpha^2+P_\alpha^2) \;\;,
\end{eqnarray}
introducing matrix notation through ($\alpha ,\beta =1,\dots ,4$):
\begin{equation}
 \hat S\equiv (S_{\alpha\beta}):=\left(
 \begin{array}{cccc}
  0&0&\sqrt{2}&0\\
   0&0&\sqrt{2}&0\\
   \sqrt{2}&\sqrt{2}&0&0\\
   0&0&0&0
 \end{array}
\right) \;\;, \;\;\;
 (E_\alpha ):=\left(
 \begin{array}{c}
  1\\
   -1\\
   0\\
   0
 \end{array}
\right) \;\;.
\end{equation}

Having a Hamiltonian function in hand, given by $H_\Sigma=H_{cl}+H_{qm}+\mathcal{I}_{hyb}$, the equations of motion are obtained through the 
Poisson brackets with respect to all canonical coordinates and momenta ($X_\alpha ,P_\alpha ,x,p$), as usual \cite{elze1,heslot}: 
 \begin{eqnarray}
 \dot{x}=\left\{x,H_\Sigma\right\}&=&\frac{p}{m} \;\;, \label{xclmoto} 
\\ [1ex] 
\dot{p}=\left\{p,H_\Sigma\right\}&=&-m\omega^2x-\frac{\beta\Omega}{2}\sum_{\alpha=1}^4 E_\alpha (X_\alpha^2+P_\alpha^2) \;\;,\label{pclmoto0} 
\\ [1ex] 
\dot{X_\alpha}=\left\{X_\alpha,H_\Sigma\right\}&=&\frac{\omega_0}{\sqrt{2}}\sum_{\beta =1}^4 S_{\alpha\beta}P_\beta +\frac{\beta\Omega}{\sqrt2} x P_{\alpha} \;\;,\label{xqmmoto}
\\ [1ex] 
\dot{P_\alpha}=\left\{P_\alpha,H_\Sigma\right\}&=&-\frac{\omega_0}{\sqrt{2}}\sum_{\beta =1}^4 S_{\alpha\beta}X_\beta -\frac{\beta\Omega}{\sqrt2} x X_{ \alpha} \;\;,\label{pqmmoto}
 \end{eqnarray} 
see also Section\,3 for explicit applications of the Poisson brackets. 

In the following, we will employ the convenient substitution $z_\alpha :=X_\alpha+\imath P_\alpha$, $\alpha =1,\dots ,4$\,, 
which replaces eq.\,(\ref{pclmoto0}) by:
\begin{equation}\label{pclmoto1}
 \dot{p}= -m\omega^2x - \frac{\beta\Omega}{2}(z^*_1z_1-z^*_2z_2) \;\;,
\end{equation}
for example, while eq.\,(\ref{xclmoto}) is unaffected.

Furthermore, by suitable linear combinations of the equations for the quantum sector (variables $X_\alpha,P_\alpha$), 
 eqs.\,(\ref{xqmmoto})--(\ref{pqmmoto}), the above substitution yields the coupled evolution equations: 
\begin{equation}\label{kqmmoto}
 \left(
 \begin{array}{c}
  \dot z_1\\
  \dot z_2\\
  \dot z_3\\
  \dot z_4
 \end{array}
\right)=\imath\frac{\omega_0}{\sqrt{2}}
\left(
\begin{array}{cccc}
 0&0&1&0\\
   0&0&1&0\\
   1&1&0&0\\
   0&0&0&0
\end{array}
\right)\left(
 \begin{array}{c}
  z_1\\
  z_2\\
  z_3\\
  z_4
 \end{array}
\right)+\imath \frac{\beta\Omega}{\sqrt2} x\left(
\begin{array}{cccc}
1&0&0&0\\
0&-1&0&0\\
0&0&0&0\\
0&0&0&0
\end{array}
\right)\left(
 \begin{array}{c}
  z_1\\
  z_2\\
  z_3\\
  z_4
 \end{array}
\right)
\;\;. \end{equation} 
Thus, we find that $z_4$ is presents a constant of motion. 

We remark that the formal solution for the classical sector (variables $x,p$) can be obtained by the method of Green's function. 
However, since the full dynamics of our model is not amenable to straightforward analytical techniques, we resort to numerical 
solutions of the coupled equations of motion (\ref{xclmoto})--(\ref{pqmmoto}).  

\section{Dynamics of the hybrid model}
We have solved the equations of motion numerically with the help of programs provided in \textit{Wolfram Mathematica 8.0}. 
In the following subsection, 
we present solutions of the full dynamics of the system, which is the quantum-classical hybrid analogue of a generalization of 
the {\it Rabi model} called the {\it Tavis-Cummings model} \cite{tavis0, tavis1}; cf. Refs. \cite{irish1,irish2,tavis2} for the 
all-quantum results. 

We have chosen a parameter regime in which the oscillator frequency $\omega$ is much larger than the characteristic frequency 
of the q-bit system $\omega_0$ which, in turn, is of the same order of magnitude as  the effective 
coupling $\lambda=\beta\omega$. The reason for this choice is that recent experiments have shown clear
spectroscopic evidence that a Cooper-pair box, or Josephson charge q-bit, coupled to a superconducting transmission 
line behaves analogously to an atom in a cavity (as in the Tavis-Cummings model). In these cases, the
dipole coupling between the two-level systems is $\beta\approx 10^{-3}$ \cite{strongcavity}. 

Capacitive or inductive couplings offer the possibility of still larger coupling strengths than those possible
with dipole coupling, even at large detuning between the fundamental frequencies of the oscillator and q-bit
system. Some results from a flux-based, inductively coupled
system give preliminary evidence for coupled quantum behaviour and entanglement between the two-level system and
the oscillator  \cite{ultracoupling}. 

In Figures\,1,\,2, we present various examples of the numerical solutions obtained for the variables $z_\alpha ,\;\alpha =1,\dots ,4$\,, 
which characterize the quantum sector of the hybrid model; we recall that $z_4$ remains constant in time and, therefore, is 
not represented here. Various combinations of initial q-bit states and model parameters are shown, with full (dashed) lines showing 
the real (imaginary) parts of the $z_\alpha$. The classical {\it oscillator 
initial conditions} always have been $x_0=0,\;p_0=1\;$; since explicit behaviour of the oscillator is not relevant for the following, we skip it here. However, various aspects of the oscillator evolution will be discussed in Sections\,4 and 5. 

Also shown in these figures is the concurrence, which will be related to a measure of entanglement of the 
two q-bits next. -- We considered a large set of initial conditions and always have found similarly 
complex behaviour of the numerical solutions, no matter whether the initial q-bit states are entangled or not.  

\subsection{Entanglement} 
For the following, we have constructed the {\it density matrix} of 
the pure state q-bit subsystem as $\hat\rho\equiv |\psi\rangle\langle\psi |$, with $|\psi\rangle$ obtained 
by employing eq.\,(\ref{oscexp}) and $X_\alpha =\mbox{Re}(z_\alpha )$, $P_\alpha =\mbox{Im}(z_\alpha )$, based on  
the numerical solutions of the full set of equations of motion, for the $z_\alpha$ and $x,p$, in particular.  

With the two-q-bit density matrix as input, we calculate the {\it concurrence} as a measure of their {\it entanglement}  
\cite{concurrence2q,concurrence}.

Let us recall the essential aspects of the notion of concurrence needed here. -- 
Given the density matrix $\hat\rho$ of a pair of quantum systems $A$ and $B$, 
we define the {\it entanglement (of formation)} for a pure state $\psi$ of the bipartite system as the entropy 
$\mathcal{E}_{pure}$ of either of the subsystems $A,B$ \cite{entanglasentropy}:
\begin{equation}
 \mathcal{E}_{pure}(\psi ):=-\mbox{tr}(\hat\rho_A\log_2{\hat\rho_A})=-\mbox{tr}(\hat\rho_B\log_2{\hat\rho_B})
\;\;, \end{equation}
where we have introduced the {\it reduced density matrices} $\hat\rho_{A,B}$, defined by $\hat\rho_A:=\mbox{tr}_B\hat\rho$ and 
$\hat\rho_B:=\mbox{tr}_A\hat\rho$. This definition can be extended for the case of mixed states.
In this case, the entanglement of formation is the {\it average entanglement} of the pure states of a decomposition,
minimized over all decompositions of $\hat\rho$:
\begin{equation}\label{entanglementformationmixed}
\mathcal{E}_{mixed}(\hat\rho):=\min\left[\sum_i p_i \mathcal{E}_{pure}(\psi_i)\right]
\;\;. \end{equation}

The goal now is to replace the right-hand side of this formal definition by an explicit function of $\hat\rho$ that allows us to  evaluate it. 
 
We begin with the ``spin-flipped'' transform of a state that we denote by a tilde, $\ket{\tilde\psi}$. 
This transformation is the standard time reversal operation for a $\frac{1}{2}$-spin particle. 
For a pure state $\ket{\psi}$ of two q-bits, it is given by:
\begin{equation}
 \ket{\tilde\psi}=(\sigma_y^{(A)}\otimes\sigma_y^{(B)}) \ket{\psi^*} 
\;\;, \end{equation}
with $\ket{\psi^*}$ the conjugate of $\ket{\psi}$. -- For a generic state of two q-bits, the obvious generalization is:
\begin{equation}
 \tilde\rho=(\sigma_y^{(A)}\otimes\sigma_y^{(B)}) \hat\rho^* (\sigma_y^{(A)}\otimes\sigma_y^{(B)})
\;\;, \end{equation}
with $\hat\rho^*$ the conjugate of $\hat\rho$. 

Next, we define the {\it concurrence} $C_{pure}$ for a pure state $|\psi\rangle$ by:
\begin{equation}
 C_{pure}(\psi):=|\braket{\psi}{\tilde\psi}| 
\;\;. \end{equation}
The ``spin-flip'' transformation of a pure product state takes the state of each q-bit to an orthogonal state, which implies $C_{pure}(\psi)=0$. 
Instead, a generic state, as for example an entangled Bell state, can be invariant except for a global phase factor, which implies $C(\psi)=1$. 
One immediately sees that $0\leq C\leq 1$. 

Most importantly, it can be shown that eq.\,(\ref{entanglementformationmixed}), for a pure state, reduces to \cite{concurrence2q}:
\begin{equation}
\mathcal{E}_{pure}(\psi)=h\left(\frac{1+\sqrt{1-C^2}}{2}\right)=:\mathcal{E}(C)
\;\;, \end{equation}
with:
\begin{equation}
 h(x):=-x\log_2x-(1-x)\log_2(1-x).
\end{equation} 
Thus, the entanglement of formation reduces to an explicit function of the concurrence. 

Furthermore, a generalization of the concurrence for mixed states has been given in Ref.\,\cite{concurrence}:
\begin{equation}
 C_{mixed}(\rho):=\max[0,\sqrt{\lambda_1}-\sqrt{\lambda_2}-\sqrt{\lambda_3}-\sqrt{\lambda_4}],
\end{equation}
in terms of  the eigenvalues $\lambda_i,\;i=1,\dots ,4$\,, with $\lambda_i\geq\lambda_ {i+1}$, of the non-Hermitian matrix ${\hat R:=\hat\rho\tilde \rho}$, 
With this definition of the concurrence, {\it entanglement of formation} of two q-bits can be expressed similarly as above:
\begin{equation}
 \mathcal{E}_{mixed}(\rho)=\mathcal{E}(C)
\;\;, \end{equation}
namely, it becomes again a function of the concurrence. Thus, the concurrence $C$ will be considered as an indicator for entanglement in the following; the 
entanglement of formation, $\mathcal{E}(C)$, could be further evaluated, as we have discussed here.

In Figures\,1,\,2, in the lower right quadrant, respectively, is shown the concurrence $C$ for the evolving two-q-bit states under consideration. 
We find from the numerical solutions of the equations of motion that $C$ is constant. In the following Section\,3, we will demonstrate 
this result also analytically. The result of the calculations here, therefore presents a nontrivial check for the accuracy of our numerical method.  

 \begin{figure}[!htbp]
 \begin{center}
 \includegraphics[scale=0.55]{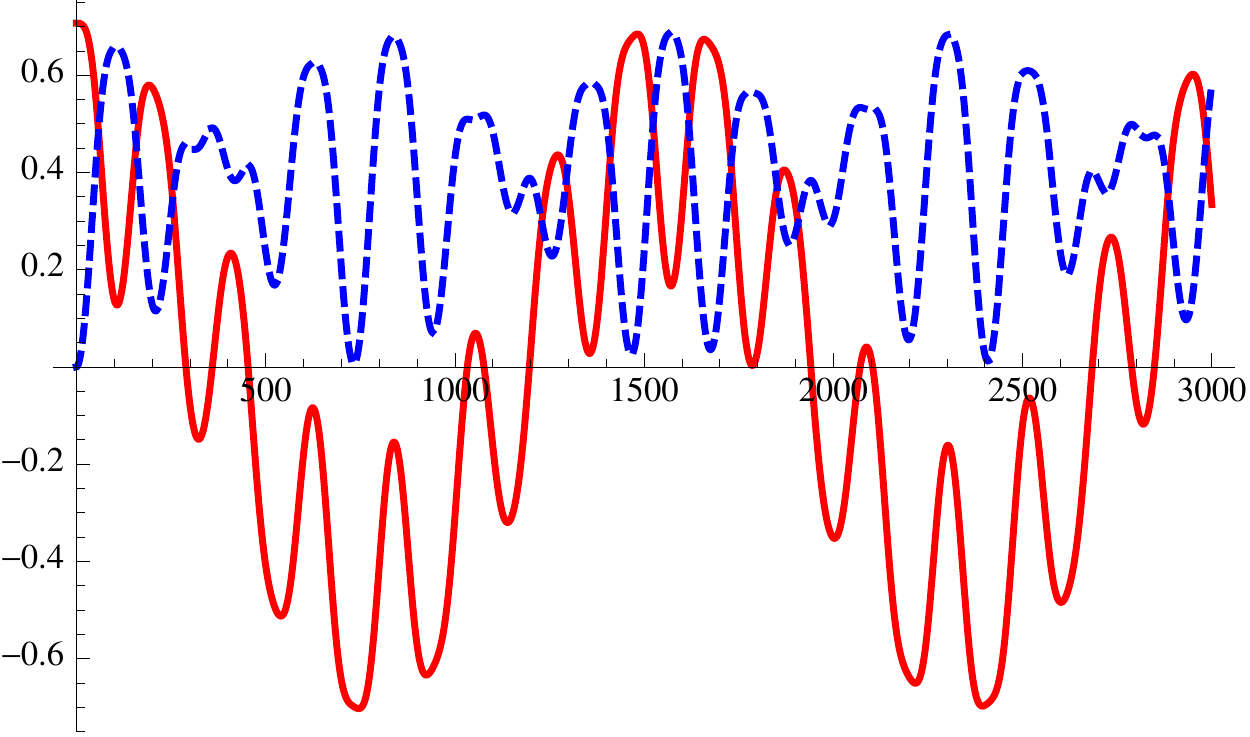}
 \includegraphics[scale=0.55]{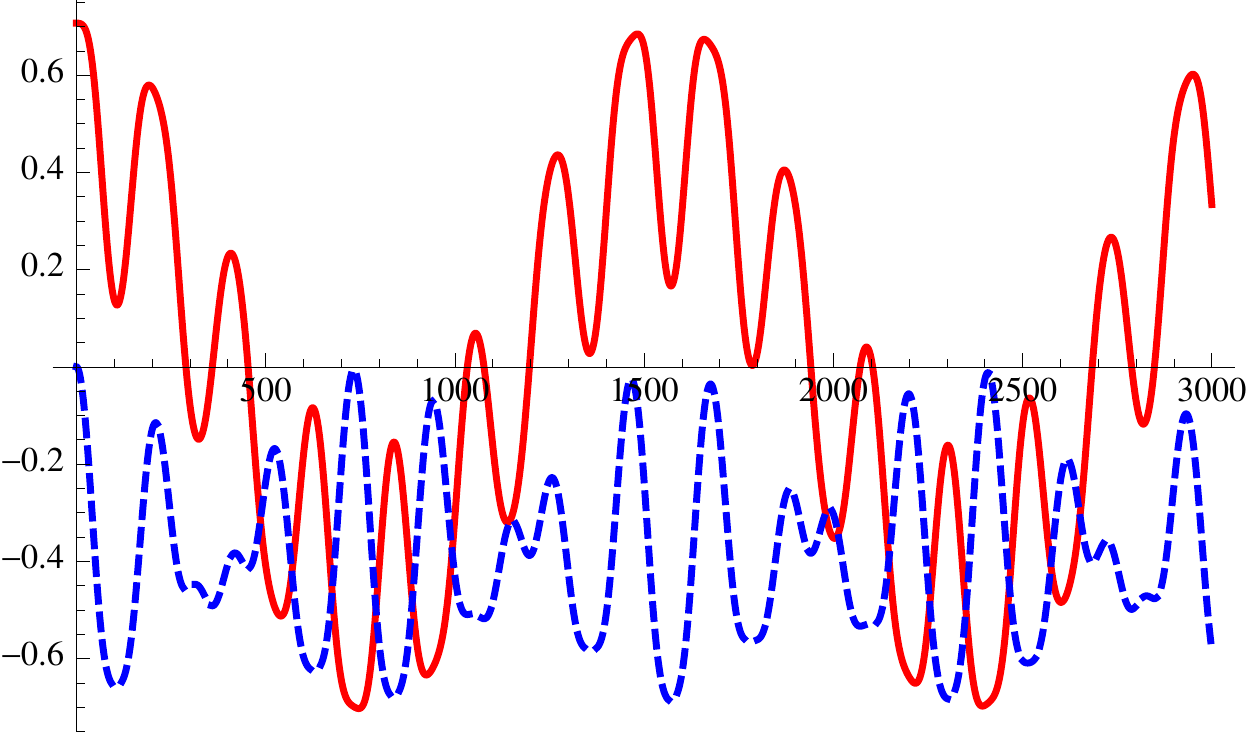}
 \includegraphics[scale=0.55]{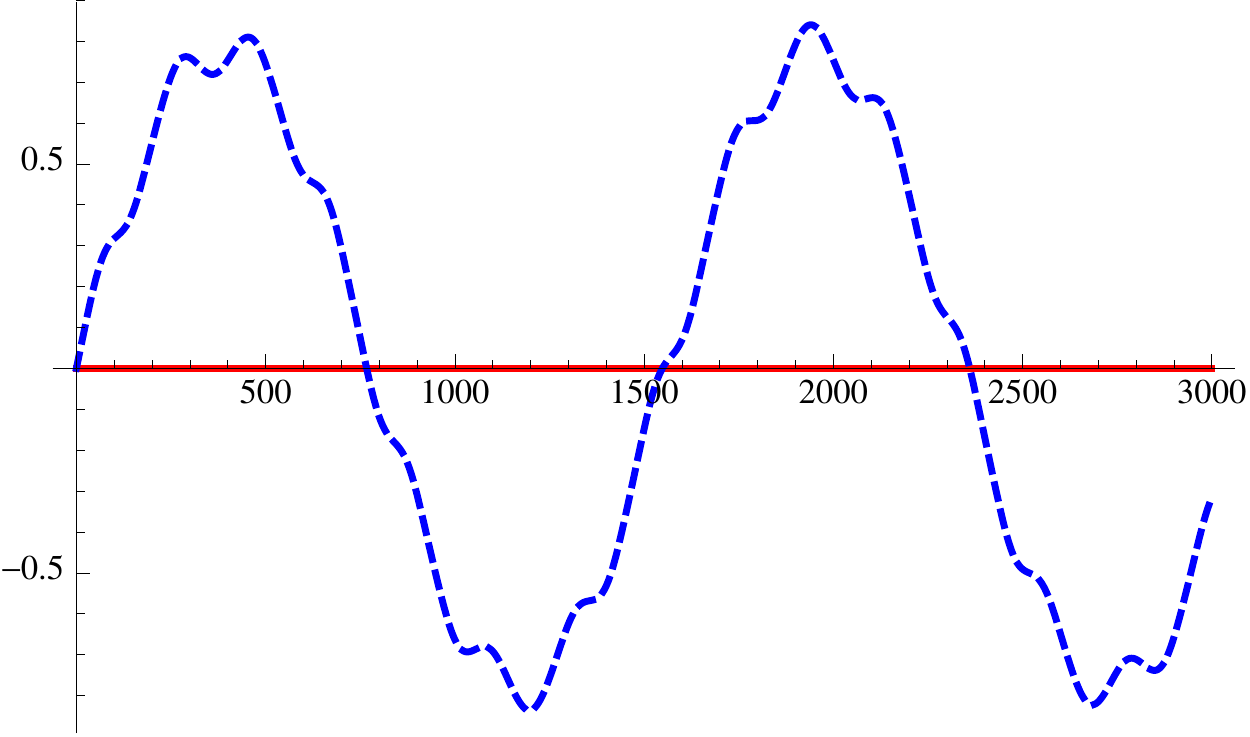} 
\includegraphics[scale=0.55]{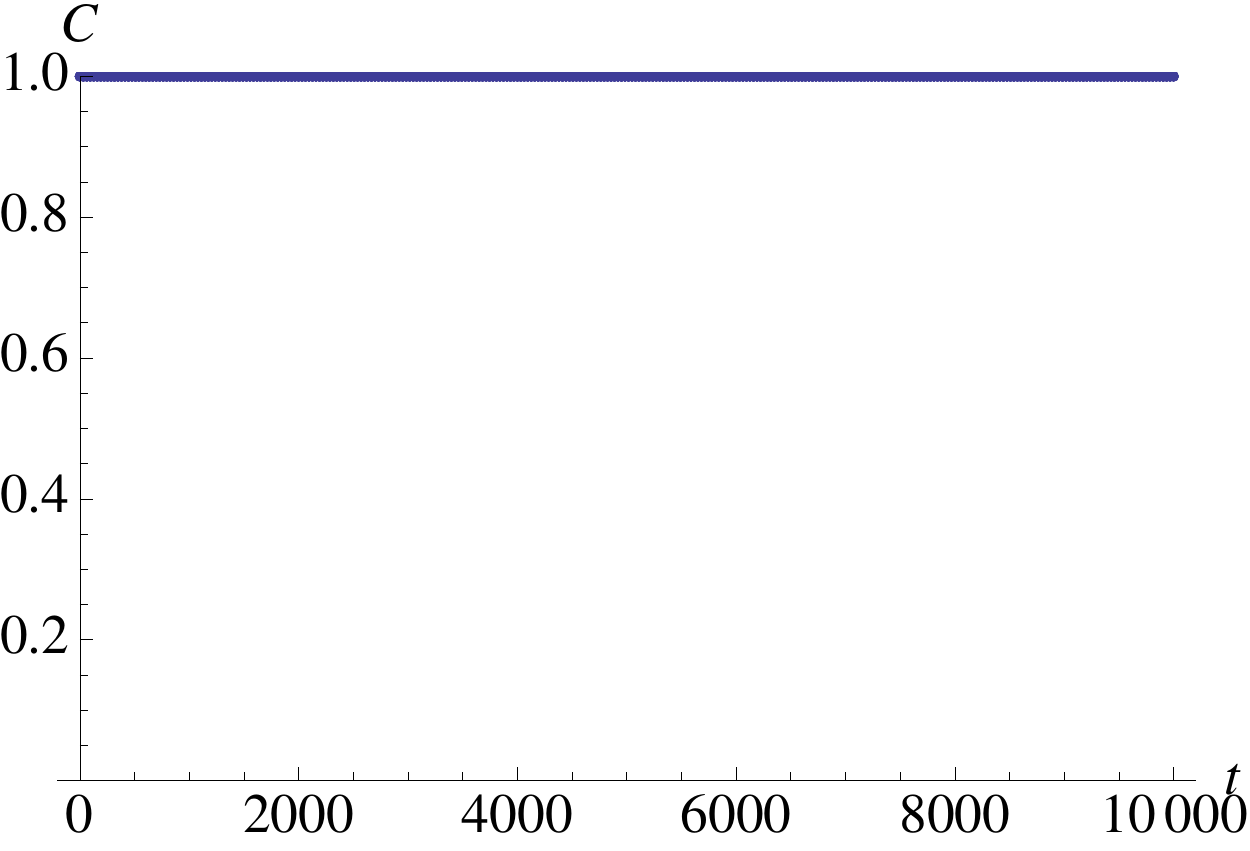} 
 \end{center}
 \caption{Evolution of $z_1(t)$ (upper left), $z_2(t)$ (upper right), $z_3(t)$ (lower left), and concurrence $C$ (lower right) 
for the initial state $\frac{1}{\sqrt{2}}(\ket{++}+\ket{--})$ 
and parameters $\omega = 0.03$, $\omega_0 = 0.15 \omega$, $\beta = 0.2$; full (dashed) lines represent real (imaginary) parts.}
\label{hyevolstates1}
\end{figure} 
 
\begin{figure}[!htbp]
\begin{center}
\includegraphics[scale=0.55]{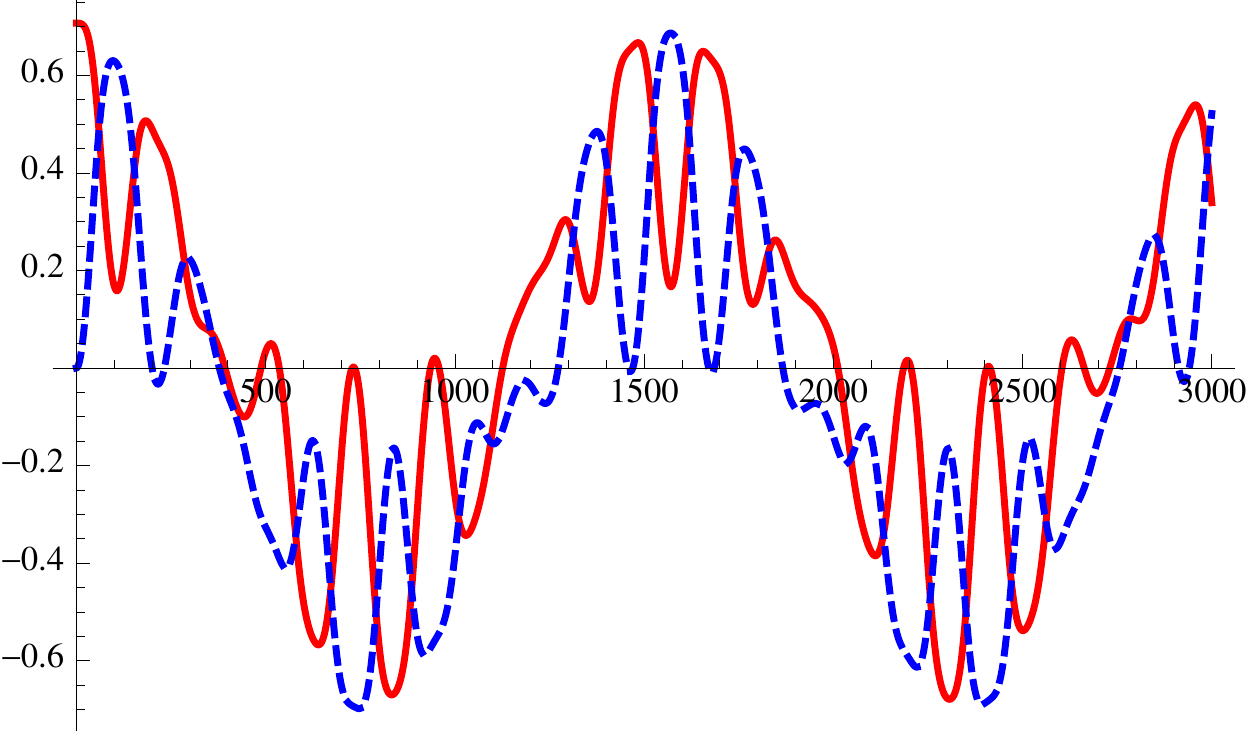}
\includegraphics[scale=0.55]{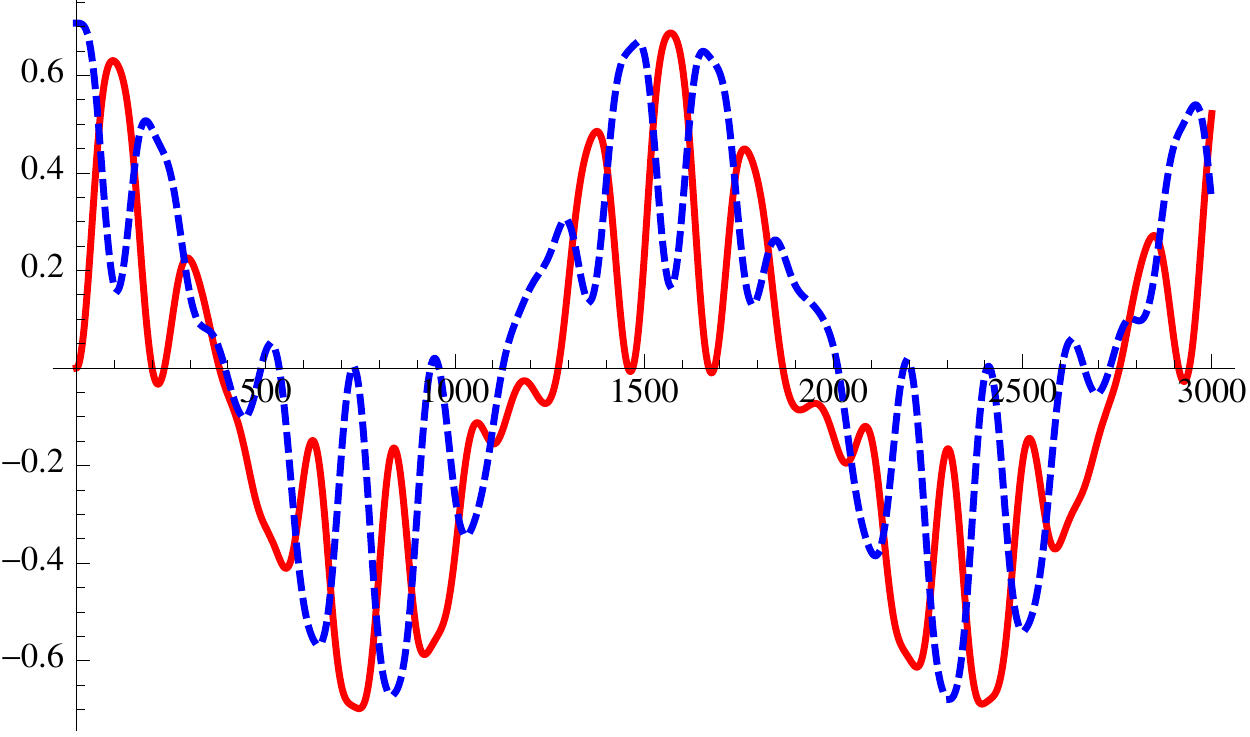}
\includegraphics[scale=0.55]{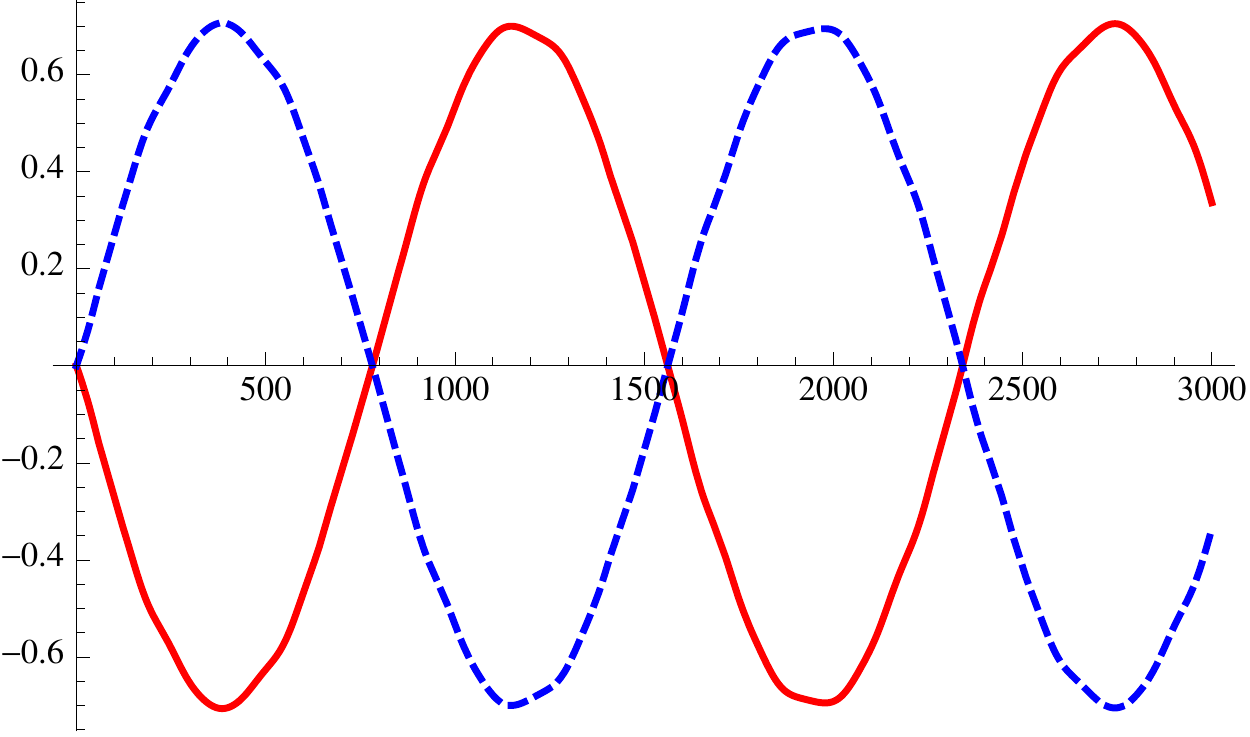} 
\includegraphics[scale=0.55]{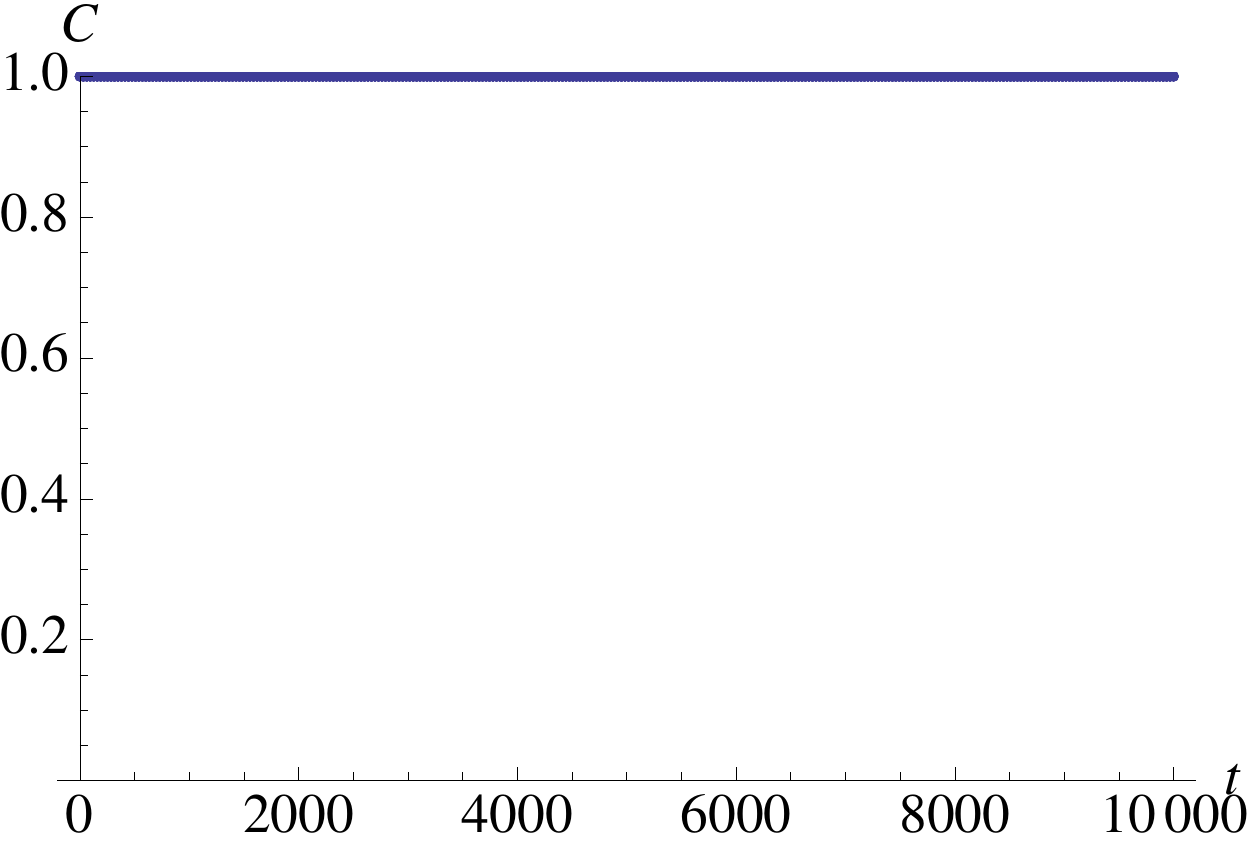}
\end{center} 
\caption{Evolution of $z_1(t)$ (upper left), $z_2(t)$ (upper right), $z_3(t)$ (lower left), and concurrence $C$ (lower right) 
for the initial state $\frac{1}{\sqrt{2}}(\ket{++}+\imath\ket{--})$  
and parameters $\omega = 0.03$, $\omega_0 = 0.15 \omega$, $\beta = 0.16$; full (dashed) lines represent real (imaginary) parts.}
\label{hyevolstates4} 
\end{figure} 

 \section{Analytical and semiclassical aspects of concurrence}\label{concost}
Instead of the numerical results presented so far, we will analyze the dynamics of entanglement in the hybrid model 
analytically here in terms of the {\it concurrence} \cite{concurrence2q}.

If the initial two-q-bit state is a generic pure state $\ket{\xi(0)}$, the oscillator expansion (\ref{oscexp}) yields:
\begin{equation}
 \ket{\xi(0)}=\sum_{\beta=1}^4\frac{X_\beta(0)+\imath P_\beta(0)}{\sqrt{2}}\ket{\phi_\beta}
\;\;. \end{equation}
The equations of motion, eqs.\,(\ref{xqmmoto})--(\ref{pqmmoto}) in particular, determine the evolution of this state. 
Correspondingly, we obtain the time dependent density matrix:
\begin{equation}
\hat\rho(t)=\sum_{\alpha,\beta=1}^4\frac{(X_\alpha (t)-\imath P_\alpha (t))(X_\beta (t)+\imath P_\beta (t))}{2}\ket{\phi_\beta}\bra{\phi_\alpha}
\;\;, \end{equation} 
the input for the evaluation of the concurrence, as described in Section\,2.1\,. 
 
Next, we need the spin-flipped version of the density matrix,  
$\tilde\rho(t)= (\sigma_y\otimes\sigma_y)\hat\rho(t)^*(\sigma_y\otimes\sigma_y)$. 
In the basis presently chosen together with the oscillator expansion, we have:
$$\sigma_y\otimes\sigma_y=\left(
\begin{array}{cccc}
 0 & -1 & 0 & 0 \\
 -1 & 0 & 0 & 0 \\
 0 & 0 & 1 & 0 \\
 0 & 0 & 0 & -1
\end{array}
\right)
\;\;,$$ 
and, consequently:
\begin{equation}\label{rhotilde}
 \tilde\rho(t)=\left(
\begin{array}{cccc}
 |z_2|^2 & z_1^*z_2 & -z_2z_3^* & z_2z_4^* \\
 z_1z_2^* & |z_1|^2 &  -z_1z_3^* &z_1z_4^* \\
  -z_2^*z_3 &  -z_1^*z_3 & |z_3|^2 & -z_3z_4^* \\
  z_2^*z_4 & z_1^*z_4 & -z_3^*z_4 & |z_4|^2
\end{array}
\right)
\;\;. \end{equation}

Finally, we have to find the square roots of the eigenvalues of the matrix ${\hat R:=\hat\rho\tilde \rho}$, in order to 
calculate the concurrence. -- 
To this end, we note that the rank of the density matrix $\hat\rho$ is equal to one, because it is
a projector for a pure state. Then, the rank of $R$ is the same as that of $\hat\rho$ and  the 
only non-trivial eigenvalue $\lambda$ is given by the trace of $\hat R$. Thus, we obtain from eq.\,(\ref{rhotilde}) 
and recalling $z_\alpha :=X_\alpha+\imath P_\alpha$, $\alpha =1,\dots ,4$\,:
\begin{eqnarray}
 \lambda&=&[2(X_1-\imath P_1)(X_2-\imath P_2)-(X_3-\imath P_3)^2+(X_4-\imath P_4)^2] \\ [1ex] 
&\;&\times [2(X_1+\imath P_1)(X_2+\imath P_2)-(X_3+\imath P_3)^2+(X_4+\imath P_4)^2] 
\;\;, \end{eqnarray}
which is real and non-negative, as it should be. 

A straightforward, even if lengthy, calculation shows that indeed, in the present case, the concurrence 
given by $C(\rho )=\sqrt\lambda$ is constant in time. We obtain explicitly, using the equations of motion 
(\ref{xqmmoto})--(\ref{pqmmoto}), 
that $\mbox{d}C(\rho )/\mbox{d}t=\mbox{d}\lambda /\mbox{d}t=0$. This confirms the results obtained in 
Section\,2 for pure initial states of the two q-bits by numerically solving the equations of motion. 

Naturally, since the equations of motion are based on the suitably generalized 
Poisson bracket algebra for quantum-classical hybrids \cite{elze1,elze2}, as we discussed, we can employ 
the Poisson bracket between the Hamiltonian, $H_\Sigma=H_{cl}+H_{qm}+\mathcal{I}_{hyb}$, and $\lambda$ directly, 
in order to demonstrate that $\lambda$ is a constant of motion. Thus, we find:
\begin{eqnarray}\nonumber
\Big\{H_\Sigma(x,p,X_\alpha,P_\alpha ),\lambda(X_\alpha,P_\alpha) \Big\}
=\Big\{ [H_{qm}(X_\alpha,P_\alpha )+\mathcal{I}_{hyb}(x,X_\alpha,P_\beta )],\lambda(X_\alpha,P_\beta )\Big\}
\\ [1ex] 
=\sum_{\gamma=1}^4\Big (\frac{\partial (H_{qm}+\mathcal{I}_{hyb})}{\partial X_\gamma}
\frac{\partial \lambda}{\partial P_\gamma}
-\frac{(\partial H_{qm}+\mathcal{I}_{hyb})}{\partial P_\gamma}\frac{\partial \lambda}{\partial X_\gamma}\Big )=0 
\;\;, \end{eqnarray}
which independently confirms the previous result. Here the classical part of the bracket 
does not contribute, since $\lambda$ does not depend on the classical variables $x,p$. 

In order to find possibly another invariant, we have evaluated also the Poisson bracket between the 
constraint $\mathcal{C}(X_\alpha,P_\alpha)$, eq.\,(\ref{constraint}), which is an invariant, and the present invariant 
$\lambda(X_\alpha,P_\beta)$. Unfortunately, since the bracket between these two quantities vanishes, 
$\{\mathcal{C},\lambda \} =0$, 
the resulting invariant is a trivial one. We have not found any other conserved quantity for our 
hybrid model. 
 
Finally, we observe that the classical equations of motion (\ref{xclmoto})--(\ref{pclmoto0}) represent a  
harmonic oscillator affected by an ``external force'', which describes the back reaction 
of the q-bits on the harmonic oscillator.
Calculating the Poisson bracket between this force:
\begin{equation}\label{forzante}
F(X_1,P_1,X_2,P_2):=\frac{\beta\Omega}{2}\left (X_1^2+P_1^2-X_2^2-P_2^2\right ) 
\;\;, \end{equation}
and the Hamiltonian, we obtain:
\begin{eqnarray}\nonumber
&\;&\Big\{H_\Sigma(x,p,X_\alpha,P_\alpha ),F(X_1,P_1,X_2,P_2)\Big\} \\  [1ex] \nonumber
&\;&\;\;\;=\;\Big\{ [H_{qm}(X_\alpha,P_\alpha )+\mathcal{I}_{hyb}(x,X_\alpha,P_\alpha )],F(X_1,P_1,X_2,P_2)\Big \}
\\ [1ex] 
&\;&\;\;\;=\;2\omega_0\beta\Omega \Big (X_3(P_2-P_1)+P_3(X_1-X_2)\Big ) 
\;\;. \end{eqnarray} 
This result, which generally does not vanish, demonstrates that the ``external force'' 
$F$, due to the genuine quantum-classical hybrid coupling in our model, is not a constant of motion, 
as could be expected. Its effect on energy conservation will be discussed in Section\,4.   

\subsection{Concurrence in presence of a semi-classical system}
Our results for the concurrence, in particular its constancy, are to some extent expected.  
In fact, if the quantum sector of our model is appropriately coupled with another quantum system, such as the 
uncoupled q-bits interacting with a quantum harmonic oscillator, like in the generalized Tavis-Cummings model  
\cite{irish1,irish2,tavis2,tavis0, tavis1}, then entanglement can be established between them. 
If we have only an interaction between a quantum and a classical subsystem, they cannot become entangled and 
the quantum state is confined to evolve in the subspace of states maintaining the initial quantum correlations.

\begin{figure}[!htbp]
  \begin{center}
  \includegraphics[scale=0.75]{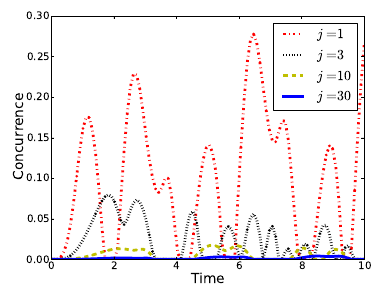}
  \includegraphics[scale=0.75]{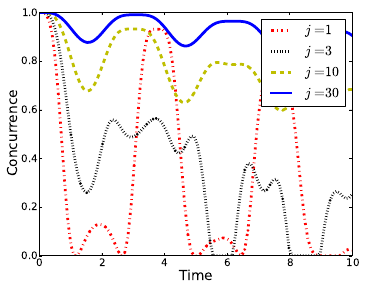}
\caption{The concurrence of two q-bits coupled by an integral spin-j
intermediary; for separable initial state $\ket{++}\ket{-j}$ of the tripartite system (left) 
and for entangled initial state $(1/\sqrt{2})(\ket{--}+\ket{++})\ket{-j}$ (right). 
Figures from \cite{polacco}.}\label{figpol1}
\end{center}
\end{figure}

\begin{figure}[!htbp]\begin{center}
  \includegraphics[scale=0.75]{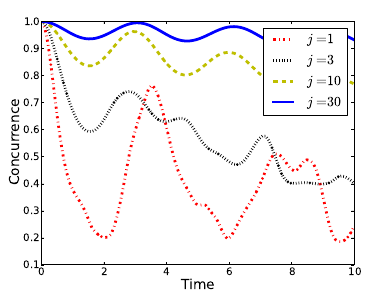}
  \includegraphics[scale=0.75]{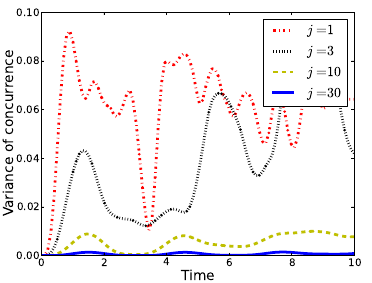}
\caption{The concurrence (left) and its variance (right) of two q-bits coupled by an integral spin-j
intermediary, representing an average over 103 initial states of the tripartite system. 
Figures from \cite{polacco}.}\label{figpol2}
\end{center} 
\end{figure}

More specifically, our study based on hybrid theory \cite{elze1}, confirms a recent analysis of two q-bits  
coupled to a large-$j$ integral spin, which latter presents a semi-classical subsystem \cite{polacco}. 
We have included figures,
from the cited article, which illustrate the trend of the concurrence in this case. 
In the limit $j\rightarrow\infty$, the integral spin subsystem becomes analogous to 
our classical  harmonic oscillator. From Figure\,\ref{figpol1} and Figure\,\ref{figpol2}, 
the trend is clear: in the classical limit, the concurrence becomes constant in time.

In Section\,5, we study further related features in our hybrid model, namely when sharp classical oscillator 
initial conditions are replaced by a phase space distribution. 

\subsection{Concurrence in presence of external perturbations}\label{concpert}
We may wonder how the hybrid system is affected, 
if we switch on an external perturbation acting on a single q-bit of this general form: 
\begin{equation}\label{perturb}
\mathcal{H}_{pert}^{(1)}:= \mathbf{I}\otimes(\omega_1\hat\sigma_x+\omega_2\hat\sigma_y+\omega_3\hat\sigma_z) 
\;\;, \end{equation} 
with $\omega_{1,2,3}$ given external parameters. 
In our chosen basis, cf. Section\,1, we have:
\begin{equation}\nonumber 
\mathbf{I}\otimes\hat\sigma_x=\left(
\begin{array}{cccc}
 1 & 0 & 0& 0 \\
 0 & -1 & 0 & 0\\
 0 & 0 & 0 & -1 \\
 0 & 0 & -1 & 0
\end{array}
\right)
\;\;,\;\;\; 
 \mathbf{I}\otimes\hat\sigma_y=\left(
\begin{array}{cccc}
 0 & 0 & \imath & \imath \\
 0 & 0 & -\imath & \imath \\
 -\imath & \imath & 0 & 0 \\
 -\imath & -\imath & 0 & 0
\end{array}
\right)
\;\;, \end{equation}
\begin{equation}
  \mathbf{I}\otimes\hat\sigma_z=\left(
\begin{array}{cccc}
 0 & 0 & 1& 1 \\
 0 & 0 & 1 & -1\\
 1 & 1 & 0 & 0 \\
 1 & -1 & 0 & 0
\end{array}
\right)
\;\;. \end{equation} 
These matrices have to be inserted in between a generic state of the two-q-bit system and its adjoint, 
in order to obtain the corresponding contribution to the hybrid Hamiltonian, cf. Section\,1. 

Then, 
incorporating the oscillator expansion for the chosen basis, we may calculate the Poisson bracket 
between this perturbation and the concurrence (for pure initial states, as before in Section\,3):
\begin{equation}
\left\{\mathcal{H}_{pert}^{(1)},\lambda\right\}=\sum_{\gamma=1}^4\Big (\frac{\partial \mathcal{H}_{pert}^{(1)}}{\partial X_\gamma}\frac{\partial \lambda}{\partial P_\gamma}-\frac{\partial \mathcal{H}_{pert}}{\partial P_\gamma}\frac{\partial \lambda}{\partial X_\gamma}\Big )=0
\;\;. \end{equation}
We note that this result is independent of $\omega_{1,2,3}$, which implies that even with this kind of
perturbation the entanglement of the q-bits does not evolve. Thus,   
our results about the evolution of entanglement hold also for {\it inhomogeneous q-bits}.   

This is obviously true for any perturbation of the classical sector as well, because the concurrence 
determined by $\lambda$ depends explicitly only on the quantum canonical variables $X_\alpha,P_\alpha$.

\begin{figure}[htpb]\begin{center}
 \includegraphics[scale=0.55]{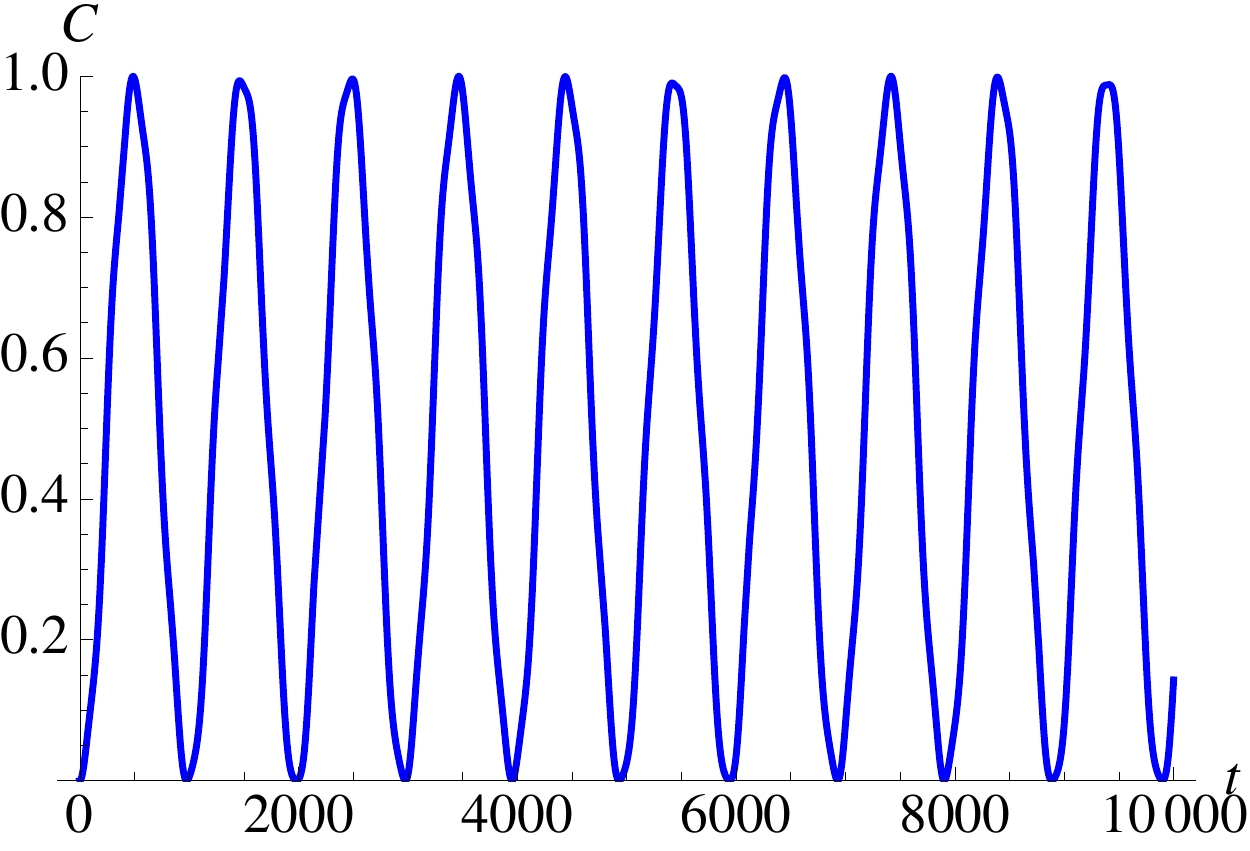}
 \includegraphics[scale=0.55]{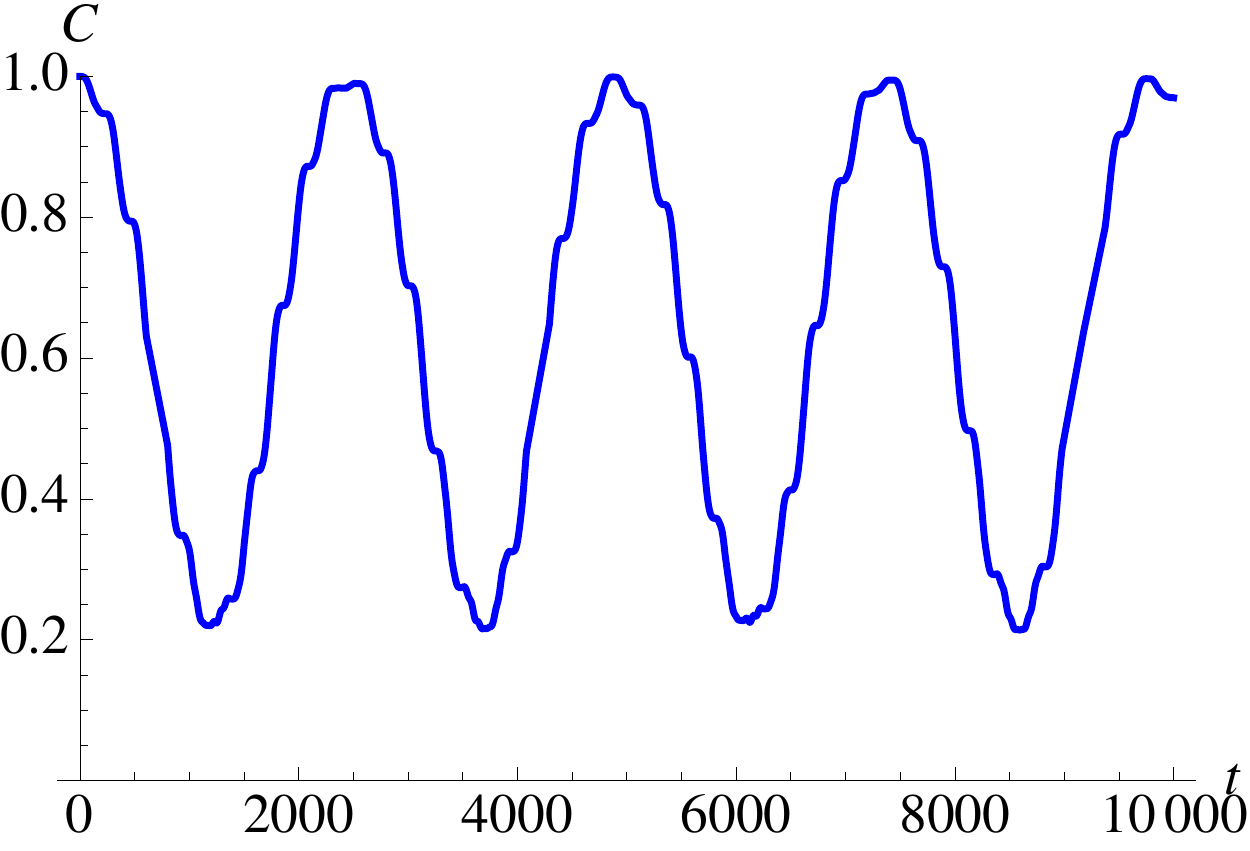}
\caption{Evolution of the concurrence in presence of a two-q-bit perturbation for the uncorrelated initial state $\ket{+-}$ (left) 
and entangled initial state $\frac{1}{\sqrt{2}}(\ket{++}+\ket{--})$ (right), with $\omega = 0.1$, $\omega_0 = 0.15 \omega$, 
$\omega_1=\omega_2=0$, $\omega_3 = 0.015 \omega$, and $\beta = 0.2$.}
\label{concurrencepert}
\end{center} \end{figure}

A different situation arises for perturbations acting on both q-bits: 
\begin{equation}\label{genperturb}
\mathcal{H}_{pert}^{(2)}:= \omega_1\hat\sigma_x\otimes\hat\sigma_x+\omega_2\hat\sigma_y\otimes\hat\sigma_y+\omega_3\hat\sigma_z\otimes\hat\sigma_z
\;\;. \end{equation} 
In the chosen basis, we obtain the corresponding matrix representation:
\begin{equation}
\hat\sigma_x\otimes\hat\sigma_x=\left(
\begin{array}{cccc}
 1 & 0 & 0& 0 \\
 0 & 1 & 0 & 0\\
 0 & 0 & -1 & 0 \\
 0 & 0 & 0 & -1
\end{array}
\right) 
\;\;,\;\;\; 
\hat\sigma_y\otimes\hat\sigma_y=\left(
\begin{array}{cccc}
 0 & -1 & 0& 0 \\
 -1 & 0 & 0 & 0\\
 0 &0 & 1 & 0 \\
 0 & 0 & 0 & -1
\end{array}
\right) 
\;\;,  \end{equation}
\begin{equation}
\hat\sigma_z\otimes\hat\sigma_z=\left(
\begin{array}{cccc}
 0 & 1 & 0 & 0 \\
 1 & 0 & 0 & 0 \\
 0 & 0 & 1 & 0 \\
 0 & 0 & 0 & -1
\end{array}
\right)
\;\;. \end{equation}
Calculating the Poisson bracket between this kind of perturbation and the concurrence, we find:
\begin{equation}
 \left\{\mathcal{H}_{pert}^{(2)},\lambda\right\}\neq 0 
\;\;. \end{equation}
This implies that the entanglement evolves in time as, for example, shown in Figure\,\ref{concurrencepert}. 
Furthermore, this result is independent of $\omega_1$, which means that perturbations of the form 
$\hat\sigma_x\otimes\hat\sigma_x$ have no effect on the entanglement, in the chosen basis. 
However, looking at the figure, we see that -- independently of whether the initial state is entangled or not -- in general, 
the two-q-bit perturbations affect the entanglement. 

We also find that the Poisson bracket between the two-q-bit perturbation and the ``external force'', eq.\,(\ref{forzante}), which describes 
the back reaction of the q-bits on the harmonic oscillator, vanishes for certain combinations of the parameters. 
We have: 
\begin{equation}
\left\{\mathcal{H}_{pert},F(X_1,P_1,X_2,P_2)\right\}=2(\omega_2-\omega_3)(X_2 P_1-X_1 P_2)
\;\;, \end{equation}
which is zero for $\omega_2=\omega_3$ and, in particular, if only $\omega_1$ is nonzero.  
In these cases, the dynamics of the quantum sector, especially of the entanglement, may change; however, the force exerted by the q-bits on 
the classical oscillator retains its form. Nevertheless, since its time dependence changes, the classical oscillator generally will be sensitive to 
the presence of perturbations acting only on the q-bits. 

\section{Quantum-classical hybrid cooling }
In this section, we would like to point out an interesting practical aspect of the quantum-classical hybrid 
coupling. Following hybrid theory, as we have described, it could be employed to transfer energy 
from the quantum to the classical sector ({\it ``hybrid cooling''}), or  vice versa. 

While the total energy of the hybrid system is a constant of motion, the energy of the classical and quantum 
sectors separately changes in time, due to the coupling. Furthermore, we have seen in Section\,\ref{concost} that the 
concurrence is conserved. This means that we can cool (or heat) a sector of the system without changing the entanglement of the 
two-q-bit states. 

\begin{figure}[!htbp]\begin{center}
  \includegraphics[scale=0.55]{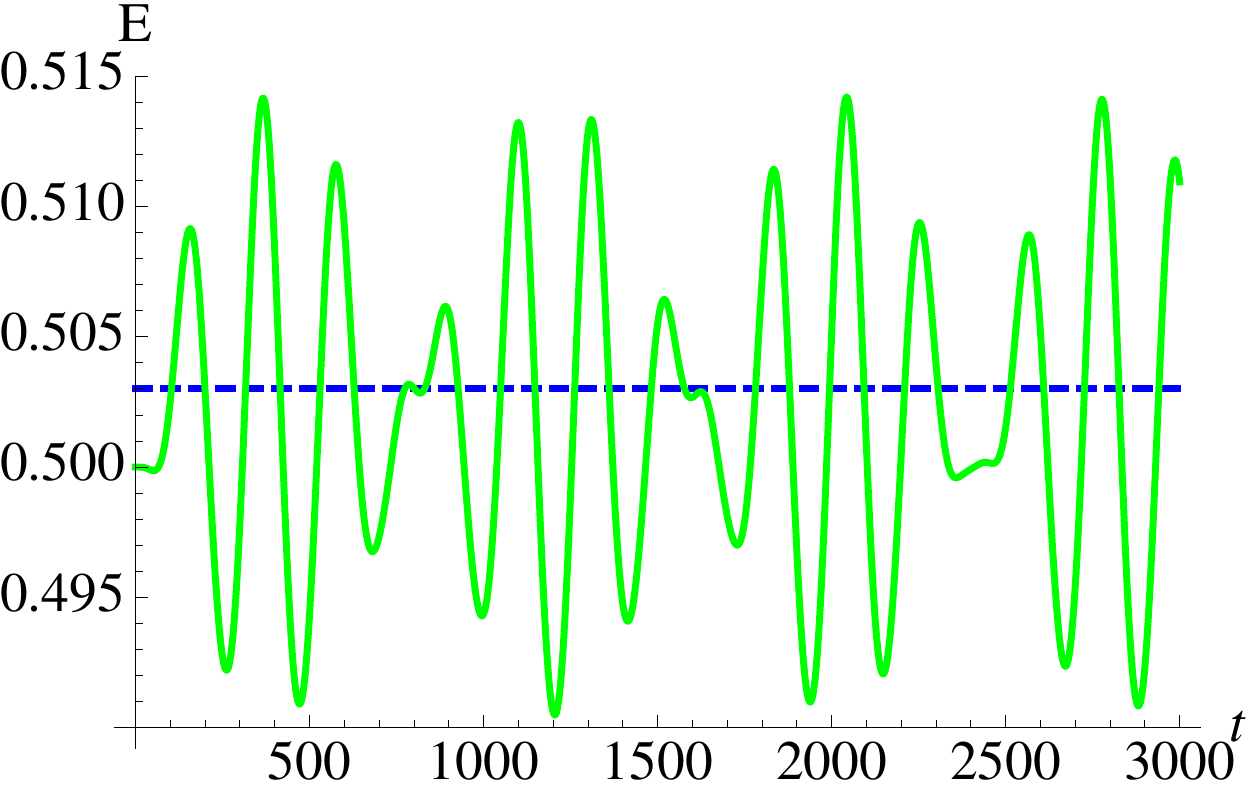}
  \includegraphics[scale=0.55]{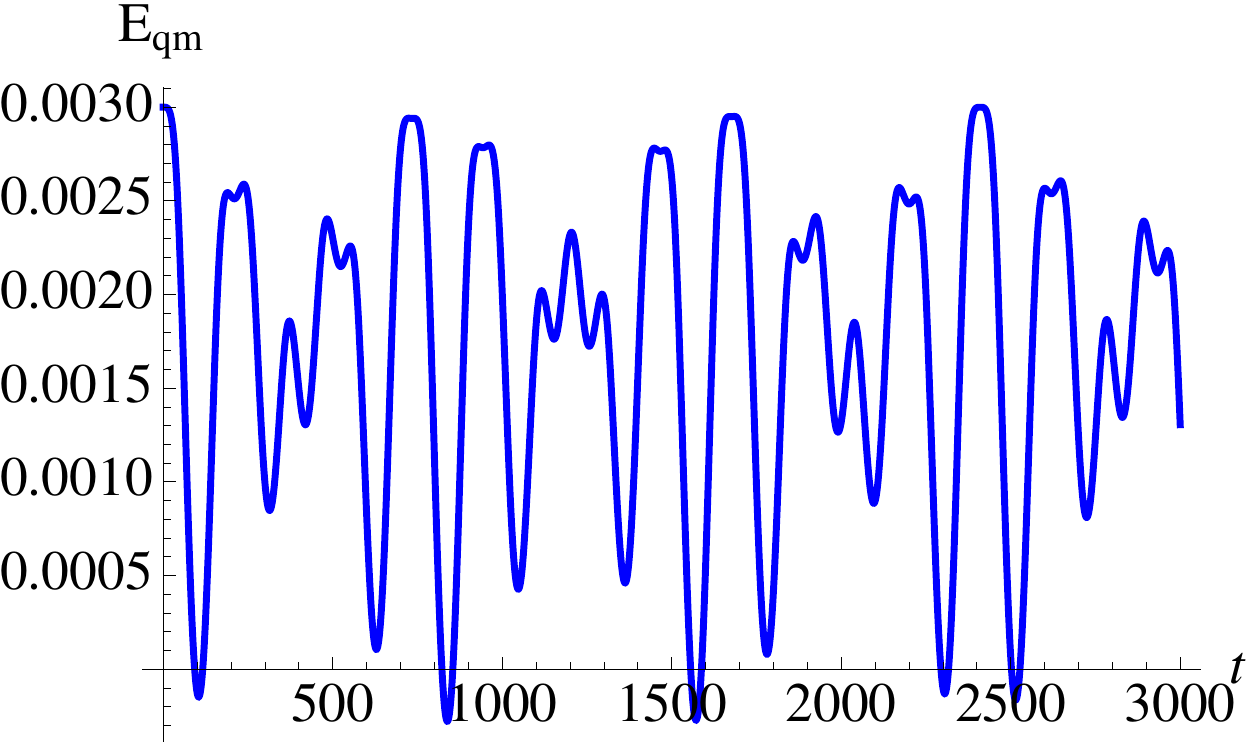}
\caption{Time dependence of energies,  for the initial state $(\ket{+-}+\ket{-+})/\sqrt 6 + (\ket{--}+\ket{++})/\sqrt 3$ and 
parameters $\omega = 0.03$, $\omega_0 = 0.16 \omega$, $\beta = 0.2$: 
on the left,  total hybrid energy dashed line (blue) and classical energy of eq.\,(\ref{energycl}) full line (green); 
on the right, quantum energy expectation of eq.\,(\ref{hqm1}) (blue).}
\label{coolingfig2}
\end{center} \end{figure}

\begin{figure}[!htbp]\begin{center}
  \includegraphics[scale=0.55]{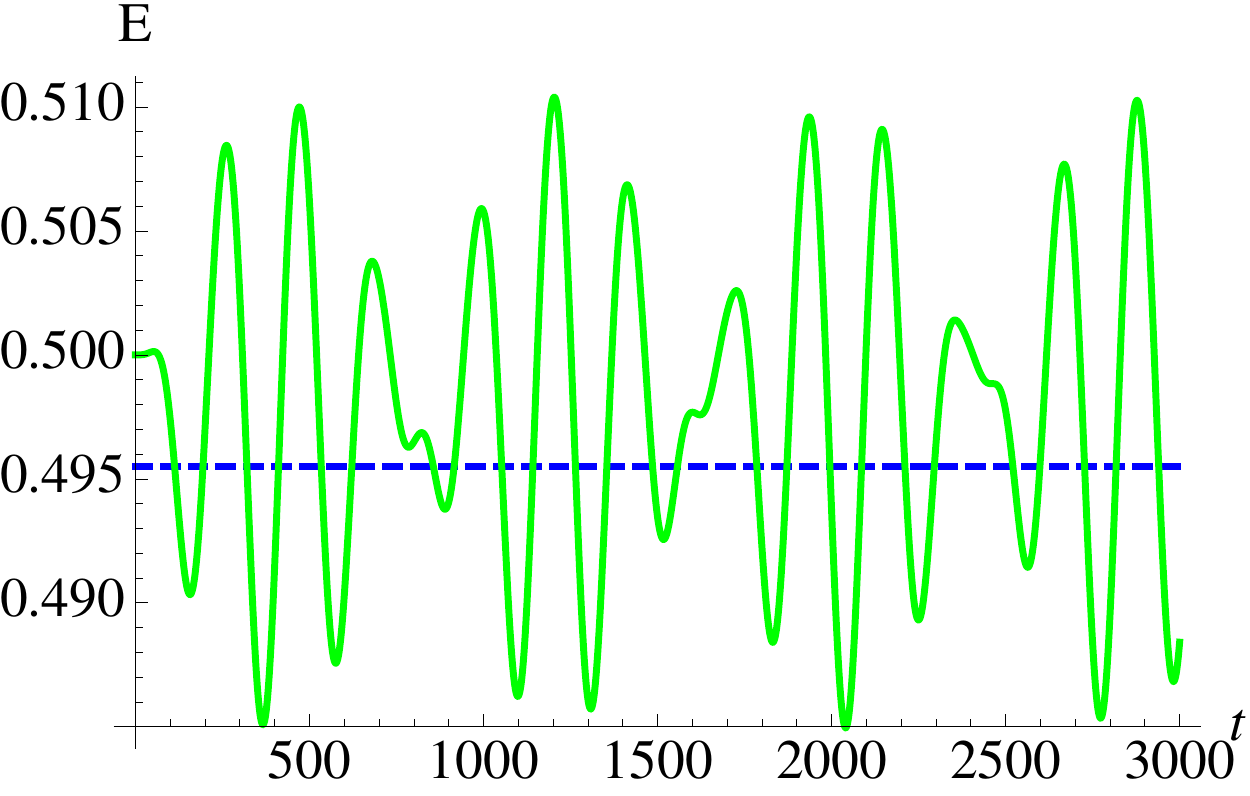}
  \includegraphics[scale=0.55]{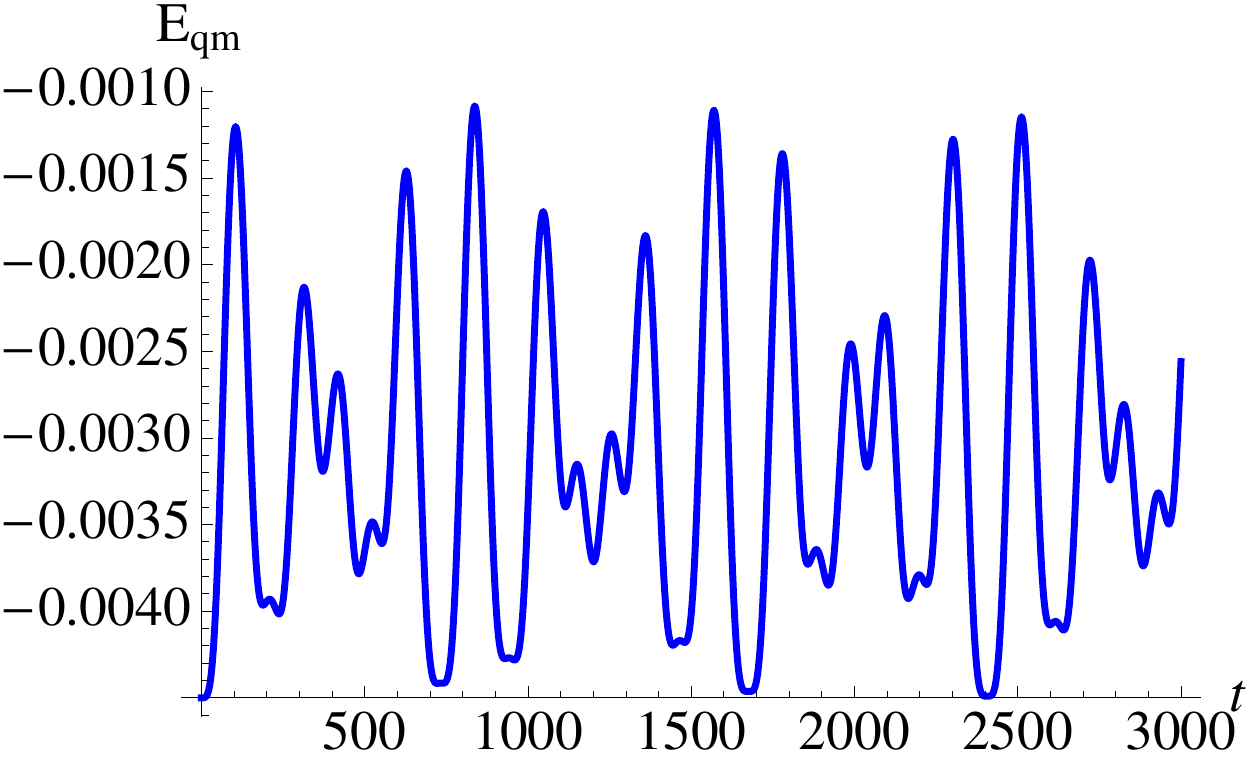}
\caption{Time dependence of energies,  for the initial state $(-\ket{++}-\ket{--}+\ket{+-}+\ket{-+})/2$ and 
parameters $\omega = 0.03$, $\omega_0 = 0.16 \omega$, $\beta = 0.02$:
on the left,  total hybrid energy dashed line (blue) and classical energy of eq.\,(\ref{energycl}) full line (green); 
on the right, quantum energy expectation of eq.\,(\ref{hqm1}) (blue).}
\label{coolingfig4}
\end{center}\end{figure}

For illustration, we have selected states that show a strong back reaction on the classical harmonic oscillator. Other 
choices are possible. Figures\,\ref{coolingfig2},\,\ref{coolingfig4} present the energy as a function of time 
for components of the quantum-classical oscillator/two-q-bits hybrid, cf. eqs. (\ref{energycl})--(\ref{hhybrid}). 
Note that the total energy is given by $H_\Sigma=H_{cl}+H_{qm}+\mathcal{I}_{hyb}$ (see Section\,1) and includes the  
contribution from the hybrid interaction (not shown in the figures). 

We see that the oscillations of the energies are wave packet like. 
The maximal amplitude of the classical energy coincides with the minimal amplitude of the quantum energy, with 
the hybrid interaction energy (not shown) carrying the remainder of the conserved total energy. 

The figures particularly suggest that -- if we can switch on the hybrid coupling for a limited period of time -- we can manipulate the energies of the 
quantum and classical sectors. This effect could have practical applications and, indirectly, test the hybrid theory. A topic for further investigations 
is how to optimize the desired control of the respective energies. 

To illustrate this effect, we choose a time-dependent coupling described by a Gaussian:
\begin{equation}\label{timedependentcoupling}
 \Omega (t)=A \exp \frac{(t-t_0)^2}{2 \sigma^2}
\;\;, \end{equation}
with the parameter $t_0$ denoting the instant of maximal hybrid coupling, $\sigma$ determining its temporal 
width and $A$ its amplitude. -- Since such coupling requires an interaction of 
the quantum-classical hybrid with yet another {\it external} system, the total energy of the hybrid is 
not necessarily conserved here.  
\begin{figure}[!htbp]\begin{center}
  \includegraphics[scale=0.5]{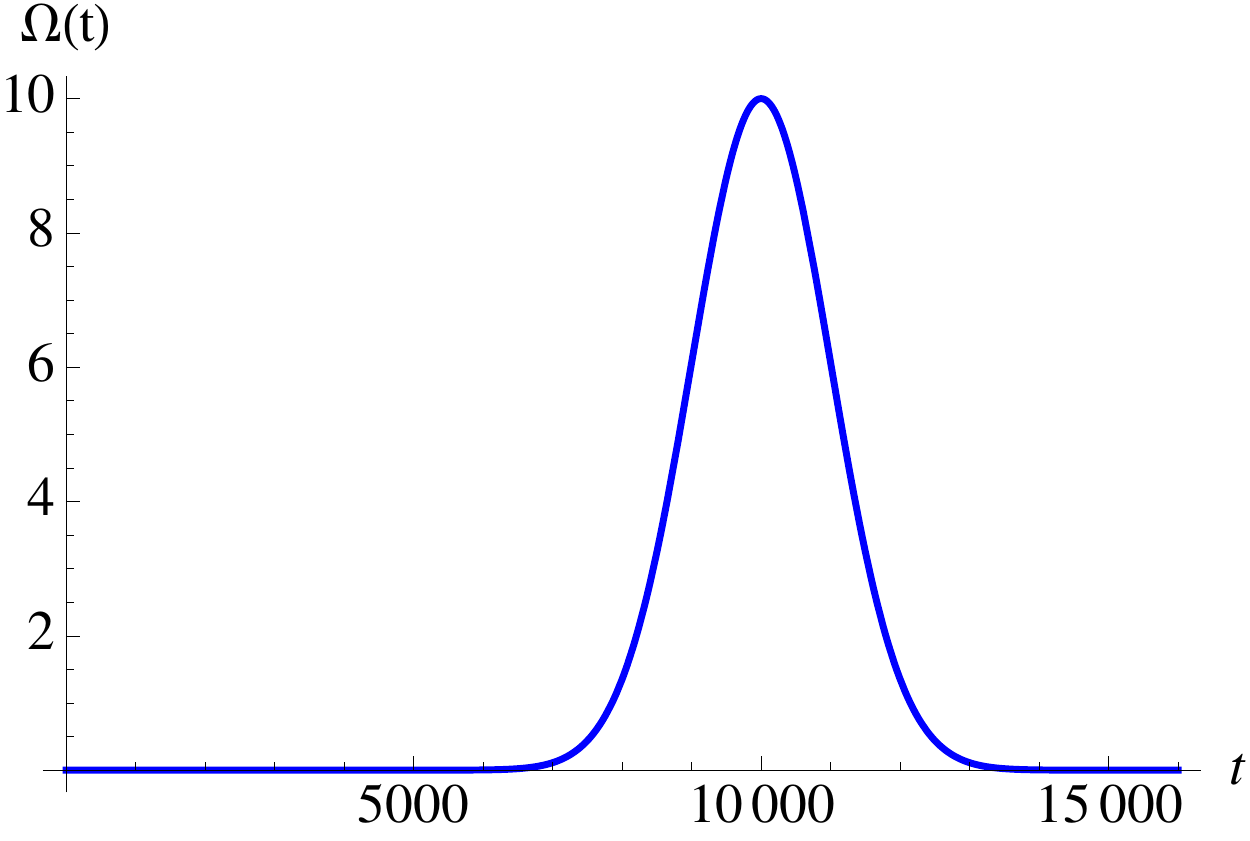}
  \includegraphics[scale=0.5]{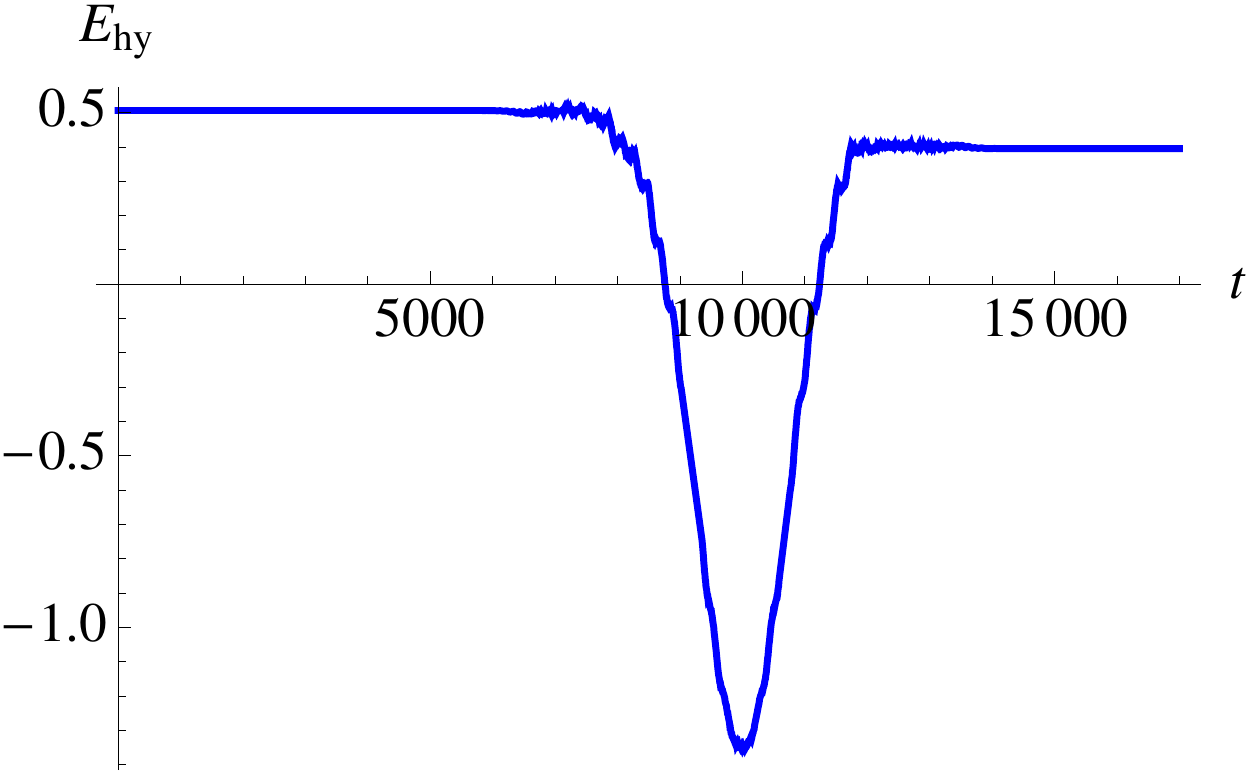}
  \includegraphics[scale=0.5]{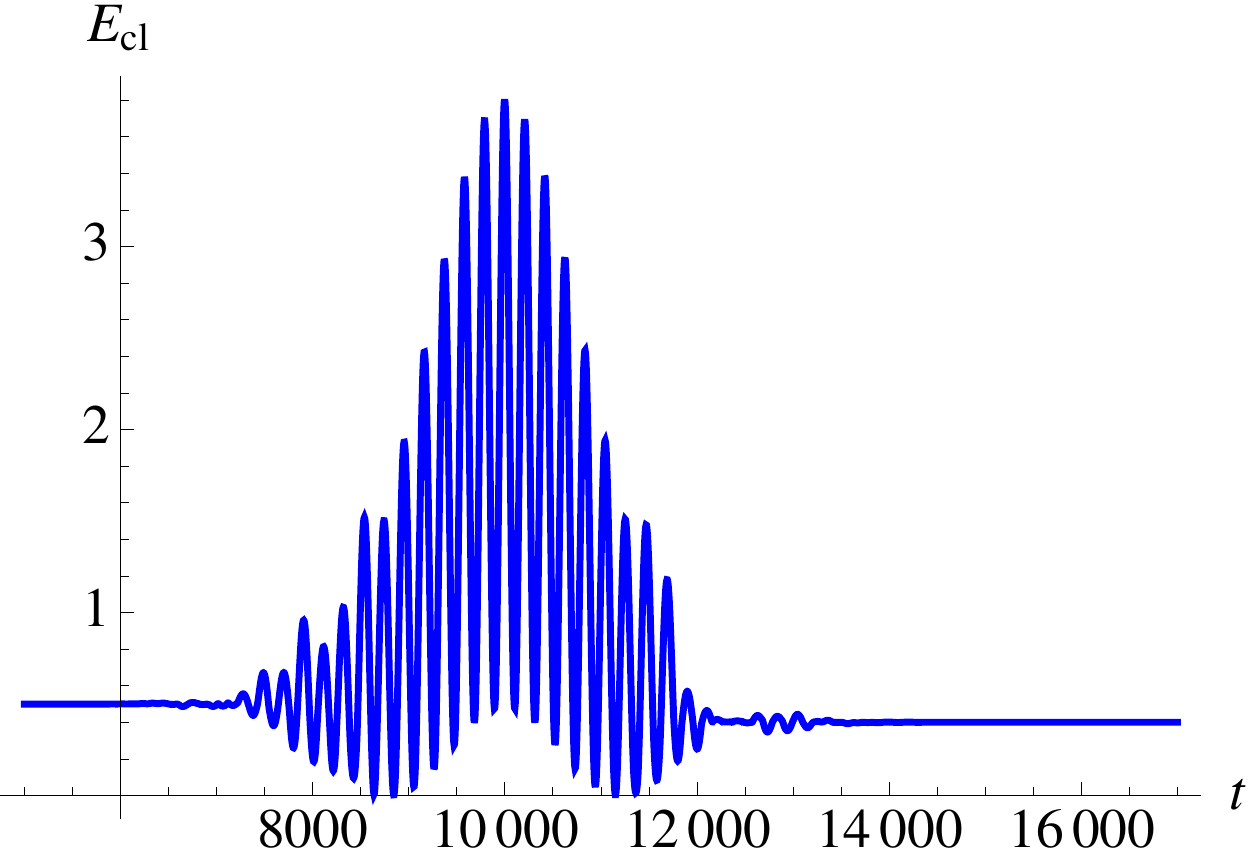}
  \includegraphics[scale=0.5]{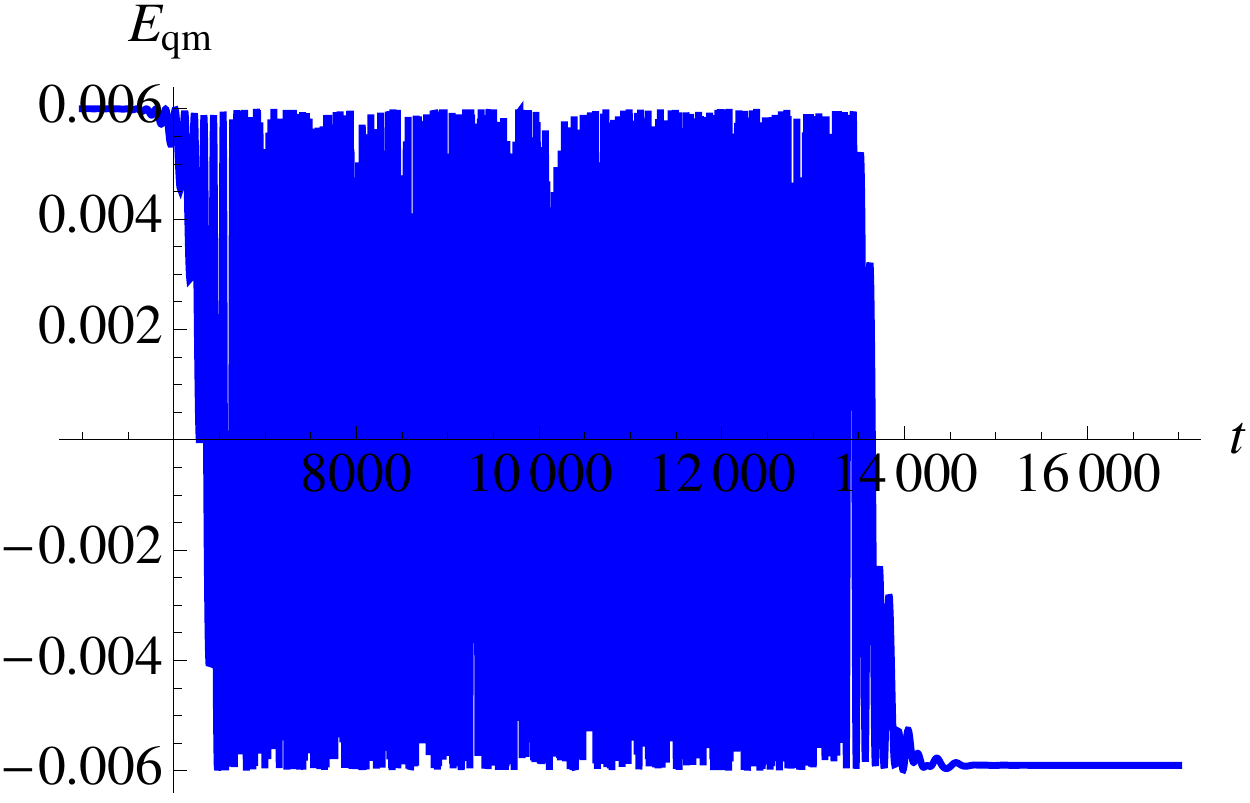}
\caption{Evolution  of energies for the time dependent hybrid coupling shown in the upper left graph  
($A=10$, $\sigma=1000$, $t_0=10^4$, cf. eq.\,(\ref{timedependentcoupling})); shown are 
the hybrid energy (top right) of eq.\,(\ref{hhybrid1}), the classical energy (bottom left) of eq.\,(\ref{energycl}),   
and the quantum energy (bottom right) of eq.\,(\ref{hqm1}). The initial state is the same as in Figure\,\ref{coolingfig2} and  
$\omega = 0.03$, $\omega_0 = 0.15 \omega$, $\beta = 0.2$.}
\label{coolingfig8}
\end{center}\end{figure} 

Depending on the choice of parameters of the hybrid coupling, the amount of cooling (and, analogously, of heating) effected on the two-q-bit subsystem 
can be varied as desired. The hybrid interaction, however, does not modify the entanglement present or absent in the initial state. 
Thus, it may offer an interesting new tool for controlling the energy of a quantum (sub)system, while leaving 
characteristic quantum features intact, as we have shown in the present example.   

\section{Quantum-classical hybrid evolution for a distribution of classical initial states}
We turn to the more realistic situation of having to deal with a distribution of initial conditions 
instead of a sharp one (i.e. a phase space point) for the classical harmonic oscillator subsystem. 
Incorporating this into hybrid initial states, we may ask how the average over such a distribution modifies 
the properties of the two-q-bit subsystem, which we have studied.  

We choose Gaussian distributions of the initial positions and momenta of the oscillator:
\begin{equation}\label{gaussianinitial}
x_0=\frac{1}{\sigma\sqrt{2\pi}}e^{(-\frac{(x-\bar x)^2}{2\sigma^2})} \;\;, \;\;\;  p_0=\frac{1}{\sigma\sqrt{2\pi}}e^{(-\frac{(p-\bar p)^2}{2\sigma^2})}
\;\;. \end{equation}

\begin{figure}[!htbp]\begin{center}
  \includegraphics[scale=0.55]{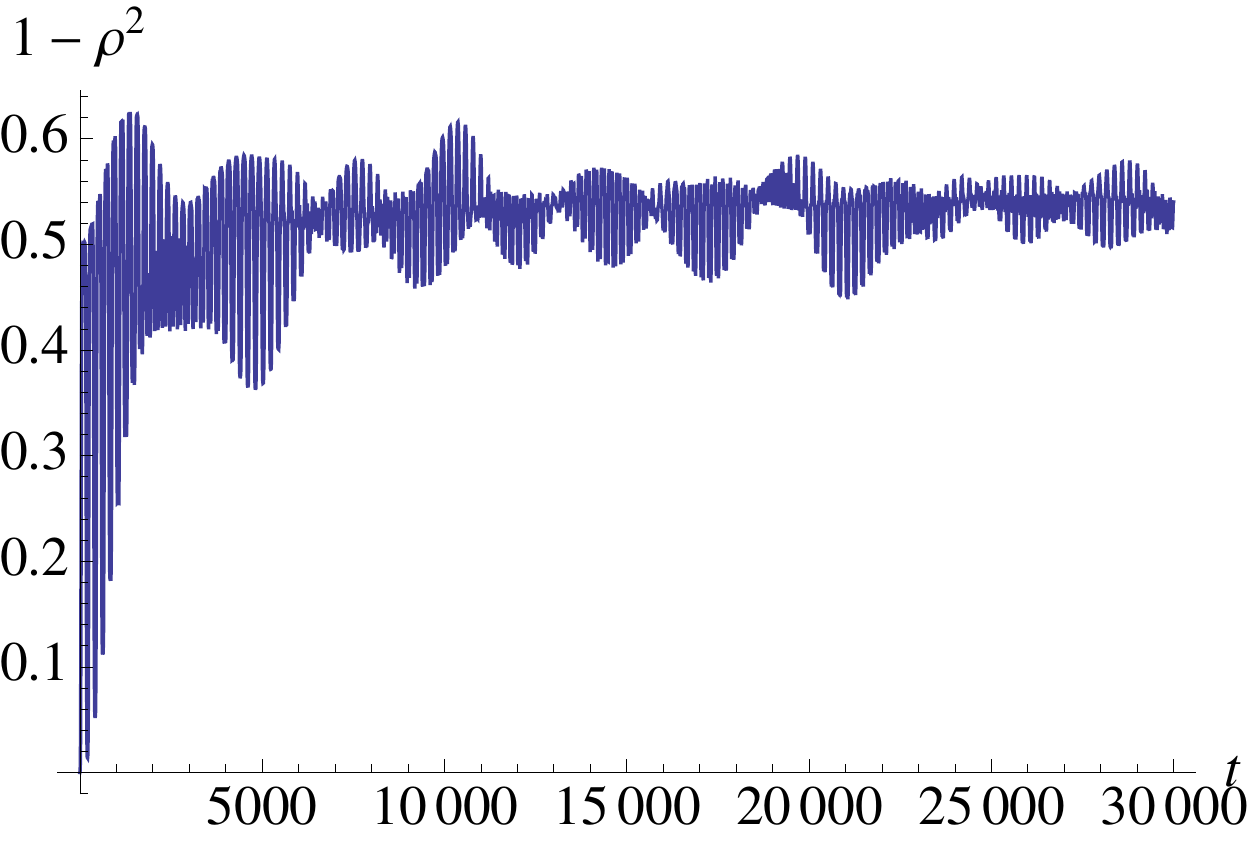}
  \includegraphics[scale=0.55]{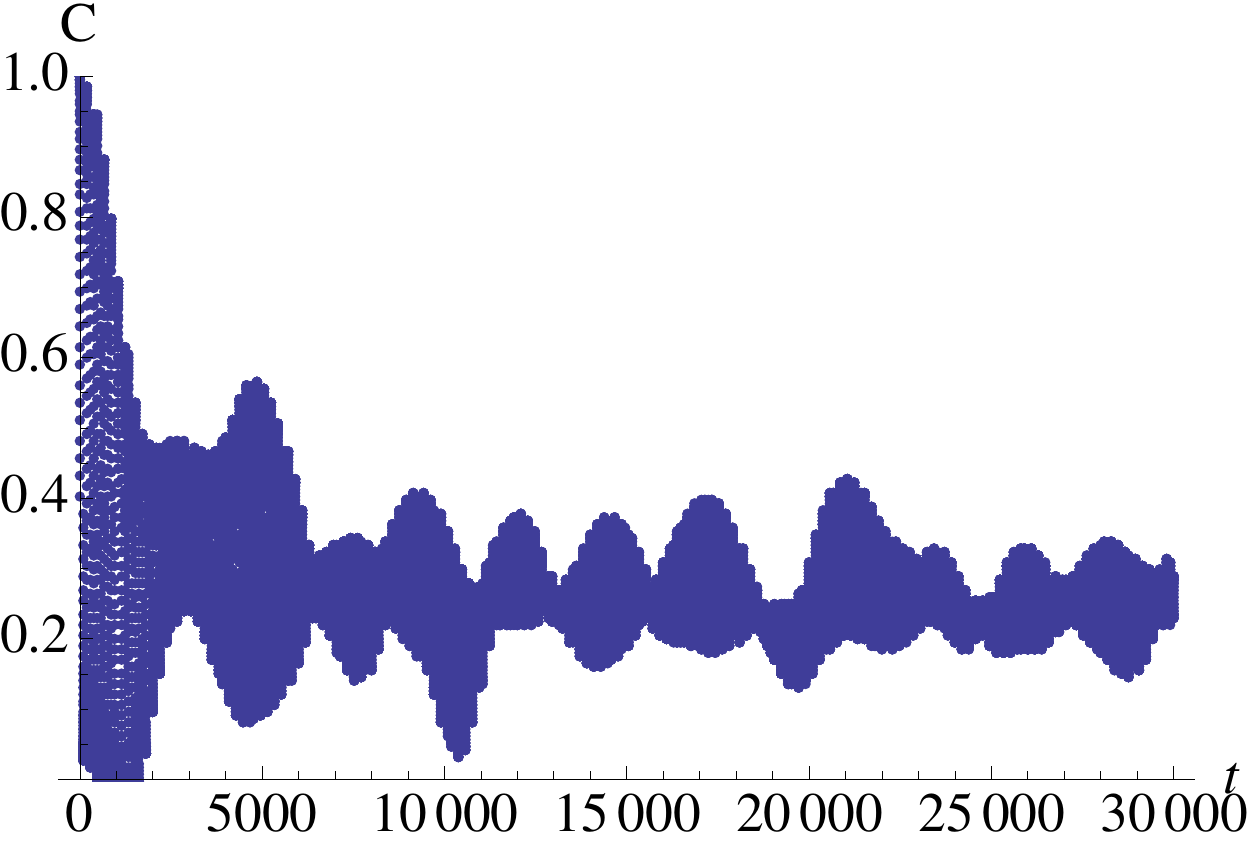}
  \includegraphics[scale=0.55]{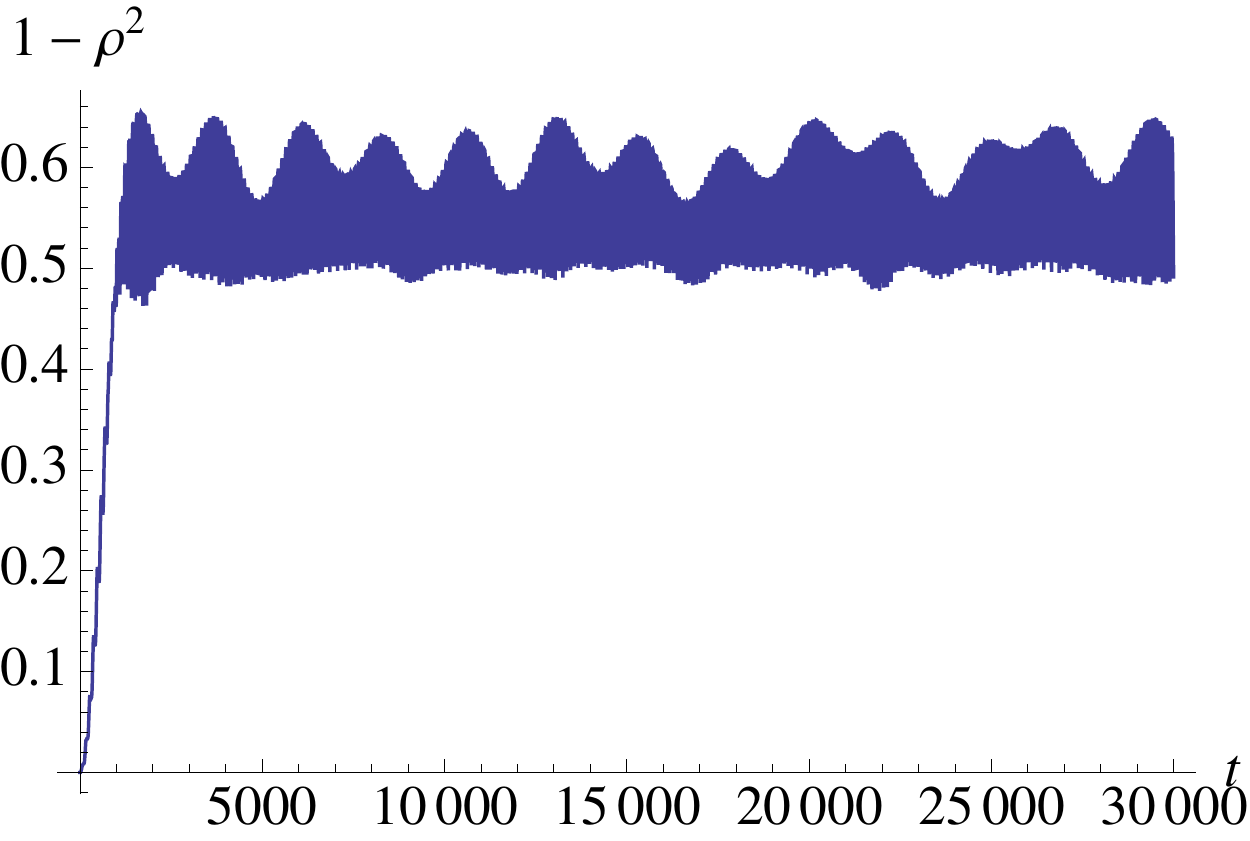}
  \includegraphics[scale=0.55]{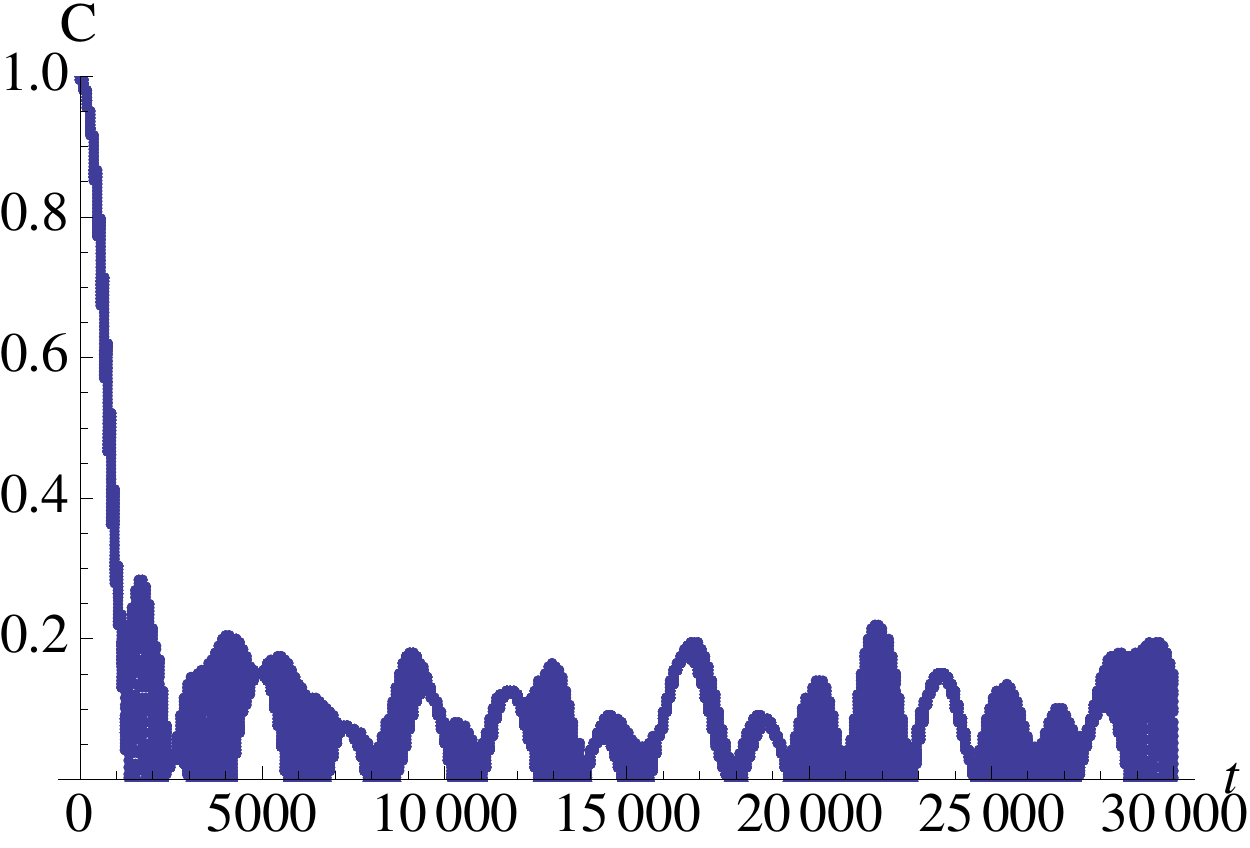}
\caption{Effects of distributed oscillator initial conditions, cf. eqs.\,(\ref{gaussianinitial}), with parameters $\bar x=0$, $\bar p=10$, $\sigma=1$. 
Shown is the evolution of the linear entropy of the two-q-bit subsystem (left figures) and of its concurrence (right figures), for couplings 
$\omega = 0.03$, $\omega_0 = 0.15 \omega$, $\beta = 0.2$;  the initial states are 
$(\ket{++}+\ket{--})/\sqrt2$ (top figures) and $(\ket{+-}+\ket{-+})/\sqrt2)$ (bottom figures).}
\label{distrclas1}
\end{center}\end{figure}

In the Figure\,\ref{distrclas1},  
we demonstrate for a typical example that a probabilistic distribution of the (classical) initial states leads effectively 
to turning the pure initial (quantum) state into a mixed one and to loosing entanglement of the two q-bits. We recall here qualitatively 
similar results obtained in Ref.\,\cite{polacco}, which we have discussed in Section\,3.1\,.

In order to obtain these results, we have used the so-called Monte-Carlo method 
with direct sampling \cite{Krauth2006}. -- 
We randomly select $M$ configurations, $\{ (x_{0,i},p_{0,i}), i=1,\dots,M\}$, 
for the initial conditions of the classical sector, which are distributed, however, 
with bias  according to eqs.\,(\ref{gaussianinitial}). Then, we  
solve the whole set of hybrid equations of motion $M$  times, once for 
each configuration $i$;  
this concerns the 
differential equations (\ref{xclmoto})--(\ref{pqmmoto}). With the help of these solutions,  
we correspondingly obtain $M$ density matrices representing the two-q-bit subsystem, 
as before. 
The average of these matrices (for $M$ sufficiently large) provides an estimate of the limiting density matrix incorporating the distributed oscillator initial conditions:
\begin{equation}\label{densexst}
\hat\rho\approx\frac{1}{M}\sum_{i=1}^M\hat\rho(x_{0,i},p_{0,i})
\;\;, \end{equation}
in accordance with the central limit theorem. 

We have performed numerical simulations for $M=10,30,50,80,100$ and observed that 
the results stabilize for $M\stackrel{>}{\sim}50$. Therefore, in 
Figure\,\ref{distrclas1}, only results for $M=100$ are presented. Following 
Ref.\,\cite{Krauth2006}, an error analysis could be performed. We leave this 
for future work, including the development of a more powerful numerical scheme.  

Having performed a series of simulations of this kind, we find that the time in which a maximally entangled state passes from the initially maximal to zero concurrence, 
is quite sensitive to the width of the Gaussian initial state distribution. Indeed, choosing $\sigma=1$, we have assumed a rather narrow initial distribution 
corresponding roughly to a phase space cell of size $\hbar$ (in physical units). To complement the present study, it will be interesting 
to carry out more extensive (and computationally more demanding) numerical simulations with wider initial phase space distributions for the classical oscillator, 
e.g., of nanometer to micrometer size and momenta corresponding to small thermal energies relevant for optomechanical experiments.     

\section{Conclusions}
We have studied in this work the {\it quantum-classical hybrid} version of the versatile {\it all-quantum model} of one oscillator coupled to 
two q-bits \cite{irish1,irish2,tavis2,tavis0,tavis1}. In the present case, the oscillator is considered as classical, while the q-bits retain their full quantum features. 
The coupling between quantum and classical sectors has been constructed following the hybrid theory developed in Refs.\,\cite{elze1,elze2,elze3}, 
a topic that has found increasing attention recently, see, for example, Refs.\,\cite{buric1,buric2,hall1,Diosi,diosi1,peres}. 

We haven chosen the hybrid theory developed in Refs.\,\cite{elze1,elze2,elze3} for the 
present work, since earlier proposals have been beset with problems, such as, for 
example, violation of energy conservation or of unitarity, which has led to various 
``no-go'' theorems. An extensive list of such issues has been discussed and overcome 
\cite{elze1,elze2,elze3}. The only other consistent formulation, as far as we 
know, is the theory developed by Buric and collaborators \cite{buric1,buric2}, the formal results of which are, however, equivalent to the hybrid theory used here. -- The 
configuration space theory of Hall and Reginatto \cite{hall1} might provide an 
alternative consistent framework for the description of quantum-classical hybrids 
under certain circumstances. 
However, it has been shown recently that it leads to nonlocal signaling \cite{HallSignal} 
and nonseparability of degrees of freedom that are classically and quantum mechanically 
separable \cite{DiLauro}, due to the intrisically nonlinear 
formulation. Since these features can interfere with entanglement, we have not 
consulted this theory here.  

The present hybrid model may be adapted to experiments, in order to test the basic tenet of hybrid dynamics, namely that 
the direct coupling of quantum mechanical to classical degrees of freedom can be consistently formulated -- be it with 
orientation towards approximation schemes for fully quantum mechanical yet complex systems or because of considerations 
of foundational issues.  One may speculate that a quantum-classical hybrid coupling could be of fundamental
character, when microscopic quantum degrees of freedom interact directly with macroscopic classical degrees of freedom, 
e.g. elementary particles, atoms, or molecules interacting with a gravitational field (if it is and remains classical) or quantum mechanical 
systems undergoing a measurement by a classical apparatus (Copenhagen interpretation).

By evaluating the concurrence, we have found that entanglement present in the two-q-bit system is invariant under evolution according to the hybrid dynamics.  
This result may have been expected but has not been studied in a consistent dynamical theory before, except Ref.\,\cite{nellowork}. 
A possible application could be in quantum information protocols transferring information coherently between q-bits, 
yet under the controlling influence of distinct classical degrees of freedom embedded in the environment. 
 
Furthermore, we have shown that the model suggests a hybrid cooling scheme, such that the energy (temperature) of the q-bit subsystem 
can be selectively lowered without affecting the quantum correlations between its components.

An important question that remains to be studied further is whether other interesting aspects of a quantum object can 
be addressed by monitoring classical degrees of freedom coupled directly to it.  The extension of our study allowing for mixed quantum states 
should be carried out and, more generally, incorporating decoherence and dissipation due to the environment. 

On the other hand, we have seen that if (for example, the initial conditions of) such classical degrees of freedom are distributed statistically, 
correspondingly averaged density (sub)matrices for the q-bits show impurity  and disentanglement increasing with time. This appears much like 
in the all-quantum version of the model and might mask desperately searched for quantum effects such as collapse or (gravitationally induced) 
spontaneous localization of the wave function, see, for example, Refs.\,\cite{nellowork,Marshall,schlosshauer} and further references therein. 

Certainly, the study of quantum-classical hybrid systems has only begun and much more can be envisaged, be it for technologically 
interesting situations or be it as far-reaching as to wonder about the role of hybrids under biologically relevant circumstances.   

\ack{We thank N.\,Buric, G.\,Cella, and A.\,Posazhennikova for discussions. 
H.-T.\,E. wishes to thank A.\,Khrennikov in particular for the 
invitation to organize a special session on quantum-classical hybrid systems 
during the Vaxjoe meeting (QTAP, June 2013) and for his kind hospitality there. 
L.F. and A.L. gratefully acknowledge support  
through Phd programs of their institutions; A.L. has been supported by      
an ERC AdG OSYRIS fellowship.}



\section*{References}
\begin{thebibliography}{99}

\bibitem{elze1} 
H.-T. Elze,
\emph{Linear dynamics of quantum-classical hybrids},
        Phys. Rev. A 85, 052109 (2012) 

\bibitem{buric1} 
M. Radonji\'{c}, S.  Prvanovi\'{c} and   N. Buri\'{c},
\emph{Hybrid quantum-classical models as constrained quantum systems},
        Phys. Rev. A 85, 064101 (2012)

\bibitem{buric2} 
N. Buri\'{c},  I. Mendas, D. B. Popovi\'{c}, M. Radonji\'{c} and S.  Prvanovi\'{c},
\emph{ Statistical ensembles in Hamiltonian formulation of hybrid quantum-classical systems}, 
        Phys. Rev. A 86, 034104 (2012)  

\bibitem{hall1} 
M.J.W. Hall and M. Reginatto,  
\emph{Interacting classical and quantum ensembles},
        Phys. Rev. A 72, 062109  (2005) 

\bibitem{Diosi}
 L. Di\'osi,  
\emph{Hybrid quantum-classical master equations}, 
        to appear in this volume (2014) [arXiv:1401.0476]   

\bibitem{elze2}
 H.-T. Elze, 
\emph{Four questions for quantum-classical hybrid theory},
        J. Phys.: Conf. Ser. 361 (2012) 012004  

\bibitem{elze3} 
 H.-T. Elze, 
\emph{Proliferation of observables and measurement in quantum-classical hybrids}, 
         Int.J.Qu.Inf.(IJQI) 10, No. 8 (2012) 1241012  

\bibitem{elze4} 
 H.-T. Elze, 
\emph{Quantum-classical hybrid dynamics - a summary}, 
          J. Phys.: Conf. Ser. 442 (2013) 012007 

\bibitem{atomcavity}
C.J. Hood, T.W. Lynn, A.C. Doherty, A.S. Parkins and H.J. Kimble,
\emph{The atom-cavity microscope: single atoms bound in orbit by single photons},
         Science 287 (2000) 1447
  
\bibitem{photocavity}
J.M. Raimond, M. Brune and S. Haroche,
\emph{Manipulating quantum entanglement with atoms and photons in a cavity},
        Rev. Mod. Phys. 73 (2001) 565

\bibitem{irish1}
E.K. Irish, J. Gea-Banacloche, I. Martin and K.C. Schwab, 
\emph{Dynamics of a two-level system strongly coupled to a high-frequency quantum oscillator},
        Phys. Rev. B 72, 195410 (2005) 

\bibitem{irish2}
E.K. Irish,
\emph{Generalized rotating-wave approximation for arbitrarily large coupling},
        Phys. Rev. Lett. 99 (2007) 173601

\bibitem{tavis2}
S. Agarwal, S.M. Hashemi Rafsanjani and J.H. Eberly,
\emph{Tavis-Cummings model beyond the rotating wave approximation: Quasidegenerate q-bits},
        Phys. Rev. A 85 (2012) 043815

\bibitem{heslot}
A. Heslot,
\emph{Quantum mechanics as a classical theory},
        Phys. Rev. D 31 (1985) 1341  

\bibitem{tavis0}
M. Tavis and F.W. Cummings,
\emph{Exact solution for an N-molecule-radiation-field Hamiltonian},
        Phys. Rev. 170 (1968) 379 
 
\bibitem{tavis1}
M. Tavis and F.W. Cummings,
\emph{Approximate solutions for an N-molecule-radiation-field Hamiltonian},
        Phys. Rev.  188 (1969) 692   

\bibitem{strongcavity}
A. Wallraff, D.I. Schuster, A. Blais, L. Frunzio, R.-S. Huang, J. Majer, S. Kumar, S.M. Girvin and R.J. Schoelkopf,
\emph{Strong coupling of a single photon to a superconducting qubit using circuit quantum electrodynamics},
        Nature (London) 431(2004) 162 

\bibitem{ultracoupling}
P. Forn-D\'{\i}az, J. Lisenfeld, D. Marcos, J.J. Garc\'{\i}a-Ripoll, E. Solano, C.J.P.M. Harmans and J.E. Mooij,
\emph{Observation of the Bloch-Siegert shift in a qubit-oscillator system in the ultrastrong coupling regime},
        Phys. Rev. Lett. 105 (2010) 237001

\bibitem{concurrence2q}
C.H. Bennett, D.P. Di Vincenzo, J.A. Smolin and W.K. Wootters, 
\emph{Entanglement of a pair of quantum bits},
        Phys. Rev. Lett. 78 (1997) 5022

\bibitem{concurrence}                         
W.K. Wootters and P. Delsing,
\emph{Entanglement of formation of an arbitrary state of two qubits},
        Phys. Rev. Lett. 80 (1998) 2245 

\bibitem{entanglasentropy}
C.H. Bennett, H.J. Bernstein, S. Popescu and B. Schumacher, 
\emph{Concentrating partial entanglement by local operations},
        Phys. Rev. A 53 (1996) 2046

\bibitem{polacco}
J. Dajka,
\emph{Disentanglement of qubits in classical limit of interaction},
       Int. J. Theor. Phys. 53 (2014) 870    

\bibitem{Krauth2006} W. Krauth, 
\emph{Statistical Mechanics: Algorithms and Computations} (Oxford Univ. Press, Oxford UK, 
2006) 

\bibitem{diosi1}
L. Di\'osi, N. Gisin and  W.T. Strunz,  
\emph{Quantum approach to coupling classical and quantum dynamics},
         Phys. Rev. A 61 (2000) 022108

\bibitem{peres}
A. Peres and D.R. Terno,  
\emph{Hybrid classical-quantum dynamics},
         Phys. Rev. A 63 (2001) 022101

\bibitem{HallSignal} 
M.J.W. Hall, M. Reginatto and C.M. Savage, 
\emph{Nonlocal signaling in the configuration space model of quantum-classical 
interactions}, 
        Phys. Rev. A 86 (2012) 054101 

\bibitem{DiLauro} F. Di Lauro, 
\emph{Mixing quantum and classical dynamics}, Tesi di Laurea Triennale 
(advisor H.-T. Elze), University of Pisa (2011), unpublished 

\bibitem{nellowork} 
A. Lampo, L. Fratino and H.-T. Elze,
\emph{The Marshall et al. optomechanical experiment in  quantum-classical hybrid theory},
        submitted to Phys. Rev. A (July 2014)

\bibitem{Marshall} 
W. Marshall, C. Simon, R. Penrose and D. Bouwmeester,
\emph{Towards quantum superpositions of a mirror},
        Phys. Rev. Lett. {\bf 91} (2003) 130401

\bibitem{schlosshauer}                         
M. Schlosshauer,
\emph{Decoherence and the quantum-to-classical transition} (Springer, Berlin, 2007) 

\end{thebibliography}
\end{document}